\documentclass[10pt,preprint]{aastex}
\usepackage{psfig}

\received{20 January 2003}  
\accepted{20 May 2003}

\newcommand{\citepeg}[1]{\citep[{e.g.,}][]{#1}}

\def\lsim{\hbox{ \rlap{\raise 0.425ex\hbox{$<$}}\lower 0.65ex\hbox{$\sim$}}}
\def\gsim{\hbox{ \rlap{\raise 0.425ex\hbox{$>$}}\lower 0.65ex\hbox{$\sim$}}}

\def\ale{\mathrel{\hbox{\rlap{\hbox{\lower4pt\hbox{$\sim$}}}\hbox{$<$}}}}
\def\age{\mathrel{\hbox{\rlap{\hbox{\lower4pt\hbox{$\sim$}}}\hbox{$>$}}}}

\slugcomment{Accepted to appear in The Astrophysical Journal (20 May 2003)}
\shortauthors{Bloom, Frail, \& Kulkarni }
\shorttitle{GRB Energetics and the GRB Hubble Diagram}
\begin{document}

\title{GRB Energetics and the GRB Hubble Diagram: Promises and
Limitations}

\author{J. S. Bloom\altaffilmark{1,2}}

\bigskip

\affil{$^1$ Harvard Society of Fellows, 78 Mount Auburn Street, Cambridge, MA 02138, USA}

\affil{$^2$ Harvard-Smithsonian Center for Astrophysics, MC 20, 60 Garden Street, Cambridge, MA 02138, USA}

\author{D. A. Frail\altaffilmark{3}}

\affil{$^3$ National Radio Astronomy Observatory, Socorro, NM 87801, USA}

\author{S. R. Kulkarni\altaffilmark{4}}

\author{$^4$ Palomar Observatory 105--24, California Institute of Technology,
            Pasadena, CA 91125, USA}

\begin{abstract}

We present a complete sample of 29 $\gamma$-ray bursts (GRBs) for
which it has been possible to determine temporal breaks (or limits)
from their afterglow light curves. We interpret these breaks within
the framework of the uniform conical jet model, incorporating
realistic estimates of the ambient density and propagating error
estimates on the measured quantities. In agreement with our previous
analysis of a smaller sample, the derived jet opening angles of those
16 bursts with redshifts result in a narrow clustering of
geometrically-corrected $\gamma$-ray energies about ${\cal E}_\gamma =
1.33 \times 10^{51}\,h_{65}^{-2}$ erg; the burst--to--burst variance
about this value is 0.35 dex, a factor of 2.2. Despite this rather
small scatter, we demonstrate in a series of GRB Hubble diagrams, that
the current sample cannot place meaningful constraints upon the
fundamental parameters of the Universe.  Indeed for GRBs to ever be
useful in cosmographic measurements we argue the necessity of two
directions. First, GRB Hubble diagrams should be based upon
fundamental physical quantities such as energy, rather than
empirically-derived and physically ill-understood distance indicators
(such as those based upon prompt burst time-profiles and spectra).
Second, a more homogeneous set should be constructed by culling
sub-classes from the larger sample.  These sub-classes, though now
first recognizable by deviant energies, ultimately must be
identifiable by properties other than those directly related to
energy. We identify a new sub-class of GRBs (``f-GRBs'') which appear
both underluminous by factors of at least 10 and exhibit a rapid
fading ($f_\nu \propto t^{-2}$) at early times ($t \ale 0.5$
day). About 10--20\% of observed long-duration bursts appear to be
f-GRBs.
\end{abstract}


\section{Introduction}
\label{sec:introduction}

The observational establishment of the extra-Galactic origin of
gamma-ray bursts (GRBs) \citep{mdk+97} enlivened the possibility of
exploiting GRBs for cosmological investigations.  The discovery of GRBs
at high redshifts (GRB\,971214, $z=3.42$, \citealt{kdr+98};
GRB\,000131, $z=4.51$, \citealt{ahp+00}) further demonstrated that
GRBs, like quasars, could be used to study the early Universe and in
particular the intergalactic medium, the interstellar medium of young
galaxies and star-formation in dusty regions (see \citealt{lb01} for a
review).

The developments in the GRB field thus appear to mirror the early
years of quasar astronomy. In both cases, astronomers were astonished
by the unexpectedly large brilliance. In both cases, the fields split
into two camps: many who used the objects for cosmology and some who
were intrigued by the workings of the central engine.

However, the large luminosity function of quasars precluded their use
as a means to determine the geometry of the Universe, i.e.  for
cosmography. The bewildering diversity of GRBs in almost any respect
-- fluence, peak luminosity -- likewise seemed to make them unsuitable
for this purpose.

Frail et al.~(2001; hereafter FKS)\nocite{fks+01} presented a complete
sample of GRBs with good afterglow follow-up and known redshifts.
Most afterglows exhibit a steepening, seen either at optical or radio
wavelengths, which is expected if the explosion geometry is conical
(opening angle, $\theta_{\rm jet}$) rather than spherical
\citep{rho97a,sph99}.  FKS interpret the breaks making several
simplifying assumptions (constant ambient density, constant
$\gamma$-ray conversion efficiency) and find the surprising result
that the seemingly most energetic bursts, $E_{\rm iso} (\gamma)$
(isotropic-equivalent $\gamma$-ray energy release), possess the
smallest beaming factor, $f_b=1-\cos(\theta_{\rm jet})$. In that
sample, the true energy release, $E_\gamma = f_b\, E_{\rm iso}
(\gamma)$, was strongly clustered.

This finding -- that GRBs are standard-energy explosions -- opens up
the possibility of using GRBs and their afterglows for
cosmography. Indeed, in the August 2001 meeting at
Garching\footnotemark\footnotetext{
\url{http://www.mpa-garching.mpg.de/$\sim$light/}} we noted that a Hubble
diagram constructed from $E_\gamma$ in the FKS sample had a smaller
scatter (0.31\,dex = 0.79\,mag) compared to the dust-corrected
$B$-band peak magnitude of Type Ia supernovae (SNe) as of 1992
($\sigma$=0.84\,mag; see Figure 3 of \citealt{vp92}). Recently, 
\citet{sch03} presented another GRB Hubble diagram based on
bursts that have two empirical distance indicators: the variability
index \citep{frr03,nmb00,rlf+01} and the lag-luminosity correlation
\citep{norr02}.  The \citeauthor{sch03} analysis find a similarly small 
scatter about a standard candle.  The current distance-indicator
sample is small (9 sources) and restricted to GRBs with known
redshifts and $\gamma$-ray observations from {\it BATSE}.

As the experience with previous efforts at cosmography has shown,
standard candles whose physics is understood (e.g., Cepheid variables)
have proved more useful and robust than those which were based on
poorly-understood or unphysical phenomena (e.g., brightest stars,
number counts).  For Type Ia SNe the physics of the explosion is
reasonably well understood. Even so, for Type Ia SNe, it is still
important to understand the intrinsic diversity (i.e., the distinction
between the so-called sub-classification of ``Branch normals'' and
``Branch peculiars''; \citealt{bfn93}), to eliminate the large
fraction ($\sim$36\%; \citealt{lft+01}) of outliers on a sound
empirical basis, and thereby increase the precision of the standard
candle. Ultimately, the distance indicator must be calibrated to high
enough accuracy to be competitive with precision cosmological
experiments such as WMAP \citep{svp+03}.

As with Type Ia SNe, the apparent constancy of the prompt energy
release in GRBs is plausible on physical grounds. Furthermore,
indications of a constant energy release come from studies of GRB
afterglows \citep{pkpp01,pk02,bkf03}, a physically distinct emission
process from the prompt burst emission.  It is reasonable to expect
that complementary afterglow data (as with photometry and spectroscopy
of Type Ia SNe) would allow us to identify potential families, or
sub-classes, in GRBs and thus compile more homogeneous samples.

With this motivation and noting that the sample of GRBs with quality
afterglow data has increased since the FKS analysis (from 17 to 29),
we revisit the current state of the GRB $E_\gamma$ Hubble diagram
paying careful attention to the error analysis and identifying
potential sub-classes of GRBs.

\section{Physical interpretations for observed temporal breaks}
\label{sec:jets}

Temporal breaks (secular steepening of a light curve) have been
observed or inferred at optical, X-ray and radio wavelengths in a
number of burst afterglows. When well-sampled broadband imaging
exists, these breaks appear to be achromatic or very nearly
achromatic. The (near) achromaticity is suggestive of a geometric
rather than an emissivity effect (such as expected from an
hydrodynamically evolving synchrotron shock). 

At present, two different geometric pictures are used to interpret the
existence of $t_{\rm break}$, the characteristic break time between
two asymptotic power-law decays (see \citealt{kg03} for a
comprehensive review). In one interpretation (the ``top hat'' model),
the energy per unit solid angle is constant within some cone with a
half-angle opening $\theta_{\rm jet}$ and falls to zero outside this
angle.
The break occurs at a time when the bulk Lorentz factor of the
shock has slowed to $\Gamma \approx 1/\theta_{\rm jet}$
\citep{rho97b,sph99}. In the other interpretation (the ``universal jet''
or ``structured jet'' model), the energy per unit solid angle varies
across the face of the jet and the observer lies at an angle
$\Theta_{\rm obs}$ from the jet axis \citep{mrw98b,zm02a}. Here, the
break occurs at a time when the bulk Lorentz factor of the shock in
the direction of the observer slows to $\Gamma(\Theta_{\rm obs})
\approx 1/\Gamma(\Theta_{\rm obs})$. The present data cannot
yet distinguish between the top-hat and universal jet models
\citep{rlr02}. The observational implication of a hybrid of the two
models has only recently been explored \citep{kg03}.

The inference of $\theta_{\rm jet}$ or $\Theta_{\rm obs}$ from $t_{\rm
break}$ requires an estimate of the evolution of $\Gamma$ as a
function of observer time. The dynamical evolution of $\Gamma$ is
theoretically prescribed by the relativistic Sedov solution
\citep{bm76}. This solution is dependent upon the isotropic-equivalent
energy in the shock and the nature of the ambient medium into which
the shock propagates for $t < t_{\rm break}$ \citepeg{spn98}.  The
isotropic-equivalent energy in the prompt burst is measured using the
observed fluence, prompt-burst spectrum, and distance to the
source. The nature of the ambient medium is generally assumed to be
homogeneous---density as a function of distance $r$ from the explosion
site as $\rho(r) = n_0 r^0$---or wind-stratified ($\rho(r) = A^*
r^{-2}$). Light curves of the more general case $\rho(r) \propto r^s$
have also been explored \citepeg{pmr98}.

Of the $\sim$30 well-studied GRB afterglow to date, there has been
only $\approx$2 bursts for which a wind-stratified media are strongly
favored statistically over homogeneous media (GRB\,980519:
\citealt{fks+00} and GRB\,011121: \citealt{pbr+02}). In the remainder,
winds are either excluded (e.g., 990510, \citealt{hbf+99,pk02}; 990123
\citealt{cl00a}) or allowed at similar significances as homogeneous
profiles (e.g., 980329; \citealt{yfh+02}; see also
\citealt{pk02}). Recently, \citet{pk03} have reanalysed the afterglow
of GRB\,990510 and showed an improved consistency with the wind
hypothesis by taking into account a structured jet.

In what follows, we construct a set of geometry-corrected energies for
GRBs, following from the formalism of FKS, under the assumption that
the early afterglow propagates into a density of $\rho(r) = n_0 r^0$
(we later address the differences when, instead, a wind-stratified
medium is assumed). We also assume a top-hat model for the structure
of the jet but note the resulting energy determinations are identical
under the universal jet model interpretation when the energy per unit
solid angle drops as $\Theta^{-2}$ (\citealt{zm02a}; see also
\citealt{kg03} and referenced therein).

\section{Formalism and Error Analysis}
\label{sec:formalism}

Following the definitions and notation of FKS and \citet{bfs01}, the
value for $E_\gamma$, the total prompt energy release in a certain
bandpass, may be found from:
\begin{equation}
E_\gamma = S_\gamma \frac{4 \pi D_l^2}{(1+z)}\, k\, f_b
\label{eq:egamma}
\end{equation}
where $S_\gamma$ is the fluence received in some observed bandpass and
$D_l$ is the luminosity distance at redshift $z$. The quantity $k$ is
a multiplicative correction of order unity relating the observed
bandpass to the standard restframe-bandpass, and the opening angle
$\theta_{\rm jet}$ is a function of the jet break time $t_{\rm break}
\equiv t_{\rm jet}$, $E_{\rm iso}(\gamma)$, and the ambient number density
$n$ (see equation~1 of FKS). Since $\theta_{\rm jet}$ is implicitly a
function of $D_l$, $z$, $k$, and $S_\gamma$, the value for $E_\gamma$
is a complex function of observables.

Expressing all physical quantities in cgs units and making the analogy
with astronomical magnitudes (distance modulus $\equiv 5 \log_{10} [
D_l/{10}~{\rm pc}]$), we rearrange equation \ref{eq:egamma} to give,
\begin{equation}
 {\rm DM} = -2.5 \log_{10} \left[S_\gamma \frac{k}{1+z} (1 - \cos
 \theta_{\rm jet}) 4 \pi \right] + 2.5 \log_{10} E_\gamma - 5 \log_{10}
 \frac{{\rm cm}}{10\,{\rm pc}},
\label{eq:dm}
\end{equation}
where the DM is the ``apparent GRB distance modulus.'' Assuming that
the total energy output is constant (${\cal E}_\gamma$) from burst to
burst, equation~\ref{eq:dm} becomes,
\begin{equation}
 {\rm DM} = -2.5 \log_{10} \left[S_\gamma \frac{k}{1+z} 
	(1 - \cos \theta_{\rm jet}) 4 \pi \right] + {\rm zp},
\label{eq:dm1}
\end{equation}
with the zeropoint zp $= 30.4940 + 2.5 \log_{10} ({\cal E}_\gamma/
1.5$ foe). The unit ``foe'' equals $10^{51}$ erg.

The variance of $E_\gamma$ is a direct measure of the accuracy with
which $E_\gamma$ can be used for cosmographic purposes.  The most
direct method to estimate the variance on $E_\gamma$ is to undertake a
simultaneous analysis of the afterglow data and prompt burst emission
to determine $f_b$ and $E_\gamma$.  With this direct approach we do
not have to worry about hidden correlations between different
quantities.

Another approach is to carry out a simple error propagation of
equation \ref{eq:egamma}. The approach is straightforward but does
require the assumption that there is no covariance in the {\it
measurement} of $S_\gamma$ and the inference of $\theta_{\rm jet}$. We
justify this assumption by noting that $f_b$ is obtained from
afterglow measurements whereas $S_\gamma$ derives from the prompt
emission -- two very different phenomena. There could well be global
correlations between the prompt burst emission and afterglow emission
(e.g., the brightest bursts may be associated with bright afterglow)
but we see no reason why our inference of $f_b$ should be {\it
observationally} correlated to $S_\gamma$, beyond the small
mathematical dependence of $f_b$ upon $E_\gamma$; see equation 1 of
\citealt{fks+01}).  Adopting this approach toward variance estimation,
we find the fractional error on the $E_\gamma$ measurement from,
\begin{eqnarray}
\left(\frac{\sigma_{E_\gamma}}{E_\gamma}\right)^2 
   &=& \left(C_{\theta_{\rm jet}} + 1\right)\left[ 
     \left(\frac{\sigma_{S_\gamma}}{S_\gamma}\right)^2 + 
     \left(\frac{\sigma_{k}}{k}\right)^2\right] + 
   C_{\theta_{\rm jet}} \left[  
     \left(\frac{\sigma_{n}}{n}\right)^2 + 
     \left(\frac{3\, \sigma_{t_{\rm jet}}}{t_{\rm jet}}\right)^2\right], 
\label{eq:fracerror}
\end{eqnarray}
where,
\begin{equation}
C_{\theta_{\rm jet}} = \left[\frac{\theta_{\rm jet}\, \sin \theta_{\rm jet}}
                      {8 \left(1 - \cos \theta_{\rm jet}\right)}\right]^2.
\end{equation}
The error in the $k$-correction ($\sigma_k$) is given by equation~8 of
\citet{bfs01} and the error on the fluence ($\sigma_{S_\gamma}$) is
estimated from the prompt GRB observations. The errors on density
($\sigma_n$) and the jet-break time ($\sigma_{t_{\rm jet}}$) are found
from afterglow modeling. Again, we have assumed that the {\it
measurements} of the four observables are
uncorrelated\footnotemark\footnotetext{
   The determinations of $t_{\rm jet}$ and $n$ are inferred from
   afterglow modeling and so are correlated to some degree. However,
   from the best observed bursts often $t_{\rm jet}$ is found directly
   from optical/IR data whereas density is best constrained with radio
   measurements. Thus, we expect the correlation on $t_{\rm jet}$ and
   $n$ to be weak. \citet{bfs01} discuss the possible correlations
   between $S_\gamma$ and $k$.}.
The error on the apparent GRB DM is \hbox{2.5 $\times
\sigma_{E_\gamma}/E_\gamma$}. If $\theta_{\rm jet} \ll 1$, then
$C_{\theta_{\rm jet}} \approx 1/16$, implying that the fractional
error on $E_\gamma$ can be easily dominated by uncertainties in the
$k$-correction and $S_\gamma$.

\section{Results}
\label{sec:results}

In FKS we adopted a value of $n=0.1$ cm$^{-3}$ for the circumburst
density for all bursts.  One clear step toward improving the
$E_\gamma$ measurements is a more realistic estimate of $n$ from
broadband modeling of the afterglow light curves.  We have now
included the best density determinations that exist for about 1/3 of
the GRBs in Table 1. In the past, large systematic differences in $n$
have been derived owing to incomplete datasets and the use of
approximate relations for estimating afterglow parameters
\citep{gs02,fyb+03}. However, with more precise photometric data and
the increasing sophistication of afterglow modeling, the true
diversity of number densities is becoming a well-established
observation. Modeling yields estimates in the range 0.1 cm$^{-3}$
\lsim\ $n$ \lsim\ 30 cm$^{-3}$ and there is now little support for
extremes of either high or low density. Herein we make use of
published estimates of $n$ and adopt a (new) canonical value of 10
cm$^{-3}$ for all remaining events.

\subsection{Energetics and a Standard Energy}

In Table 2, we show the computed energetics values for $E_\gamma$ and
the associated apparent GRB distance moduli, DM, for a cosmology with
$\Omega_\Lambda = 0.7$, $\Omega_m = 0.3$, and $H_0 = 65\,h_{65}$
km\,s$^{-1}$\,Mpc$^{-1}$.  Of the 9 GRBs with measured $z$, $t_{\rm
jet}$ and $n$, the logarithmically-weighted mean total energy release
is 1.16 foe $\pm$ 0.07 dex with a median energy of 1.33 foe. We
therefore adopt a new standard energy of ${\cal E}_\gamma =
1.33\,h_{65}^{-2}$ foe with an error of $\pm$ 0.07 dex. This standard
value is dependent upon the choice of cosmology.

\begin{figure*}[hbt]
\centerline{\psfig{file=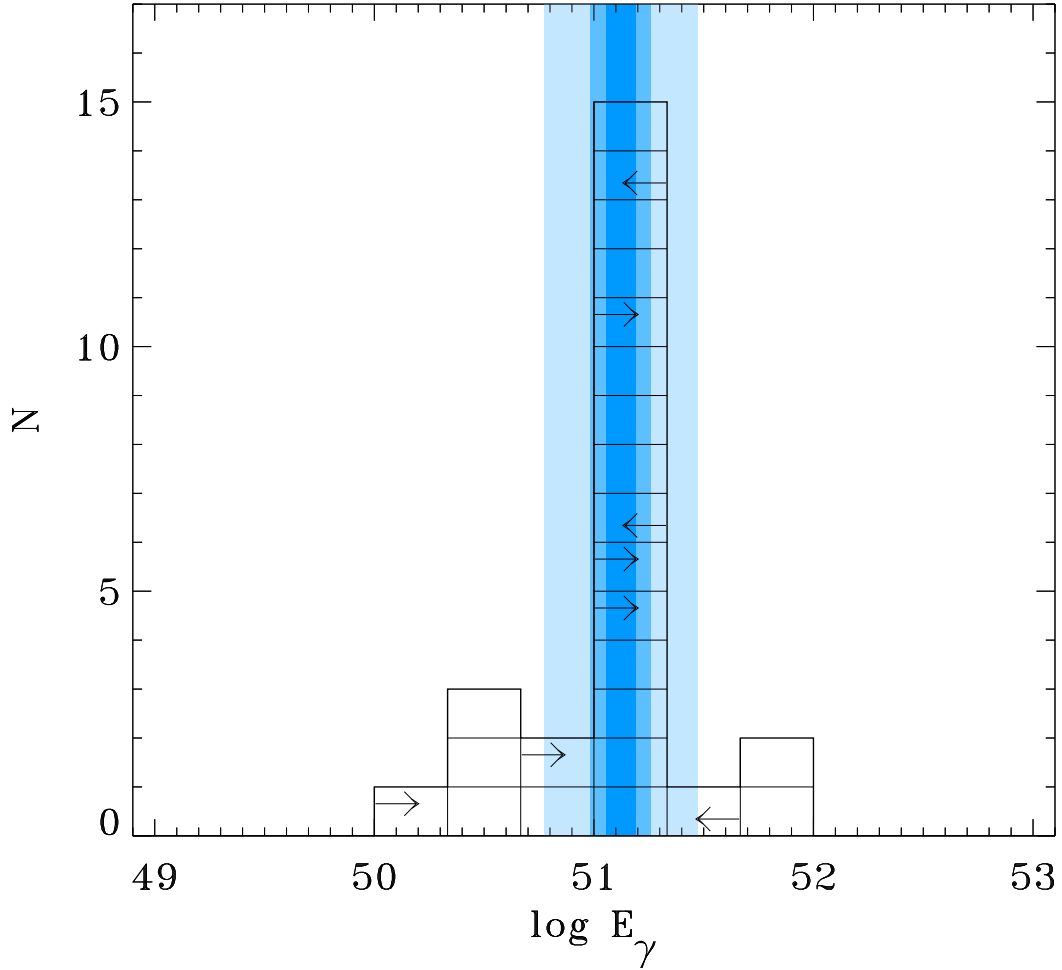,height=5.5in}}
\caption[]{A histogram of GRB energies ($E_\gamma$) with three equal 
logarithmic spacings per decade. The histogram shows a narrow
distribution of GRB energies about the standard energy ${\cal
E}_\gamma = 1.33$ foe, with an error of $\sigma = 0.07$\,dex. The
observed burst--to--burst rms spread is $0.35$\,dex (a factor of 2.23)
about this value. Bands of 1, 2, and 5 $\sigma$ about the standard
energy are shown. There are five identifiable outliers, which lie more
than 5 $\sigma$ from the mean, however, there is currently no basis
other than discrepant energy to exclude these bursts from the sample.
In Figure \ref{fig:hub2} we identify two bursts (not shown here) which
are discrepant in both energy and afterglow properties.}
\label{fig:hub3}
\end{figure*}

If we expand the sample to include all bursts with knowns redshifts
and $t_{\rm jet}$ but not density, the logarithmically-weighted mean
energy is 1.21 foe $\pm$ 0.08 dex. That these two mean values agree
suggests that the assignment of $n = 10$ cm$^{-3}$ to the 7 bursts
with only $t_{\rm jet}$ measurements reasonably reflects the density
of the population without $n$ measurements. Figure \ref{fig:hub3}
shows a histogram of $E_\gamma$ measurements, tightly clustered about
${\cal E}_\gamma$. Note that the adopted value for ${\cal E}_\gamma$,
now found using a proper inclusion of density into the $E_\gamma$
formulation, is a factor of 2.7 higher than that found in FKS. This is
close to the analytic difference expected ($\approx 3.2$ for
$\theta_{\rm jet} \ll 1$) by assuming $n= 10$ cm$^{-3}$ instead
of $n=0.1$ cm$^{-3}$, as in FKS.

As seen in Figure \ref{fig:hub3}, 11 of the 16 GRBs with $E_\gamma$
measurements fall within 5 $\sigma$ of ${\cal E}_\gamma$ and five
bursts (970508, 990123, 990510, 000418, 011211) are outside 5 $\sigma$
of ${\cal E}_\gamma$. The variance of all 16 GRBs with measured
$E_\gamma$ about our adopted standard energy is 0.35 dex and is
dominated by the five outliers. Though there is no {\em a priori}
reason to exclude these outlier bursts, the observed scatter of the
more restricted sample without the outliers about ${\cal E}_\gamma$ is
0.13 dex, about a factor of 35\%.

\subsection{$\boldmath{{\cal E}_\gamma}$ in a Cosmological Context}

The small apparent scatter of measured $E_\gamma$ values is continued
evidence for a standard energy release in prompt $\gamma$-ray
emission, first noted in FKS. As we pointed out in Garching 2001, this
observation suggests that GRBs may be used as calibrated standard
candles, useful, in principle, for measuring the geometric parameters
of the universe (``cosmography'').

\begin{figure*}[hbt]
\centerline{\large {\bf ~~~~~~~The GRB Hubble Diagram with 24 sources}}
\centerline{\psfig{file=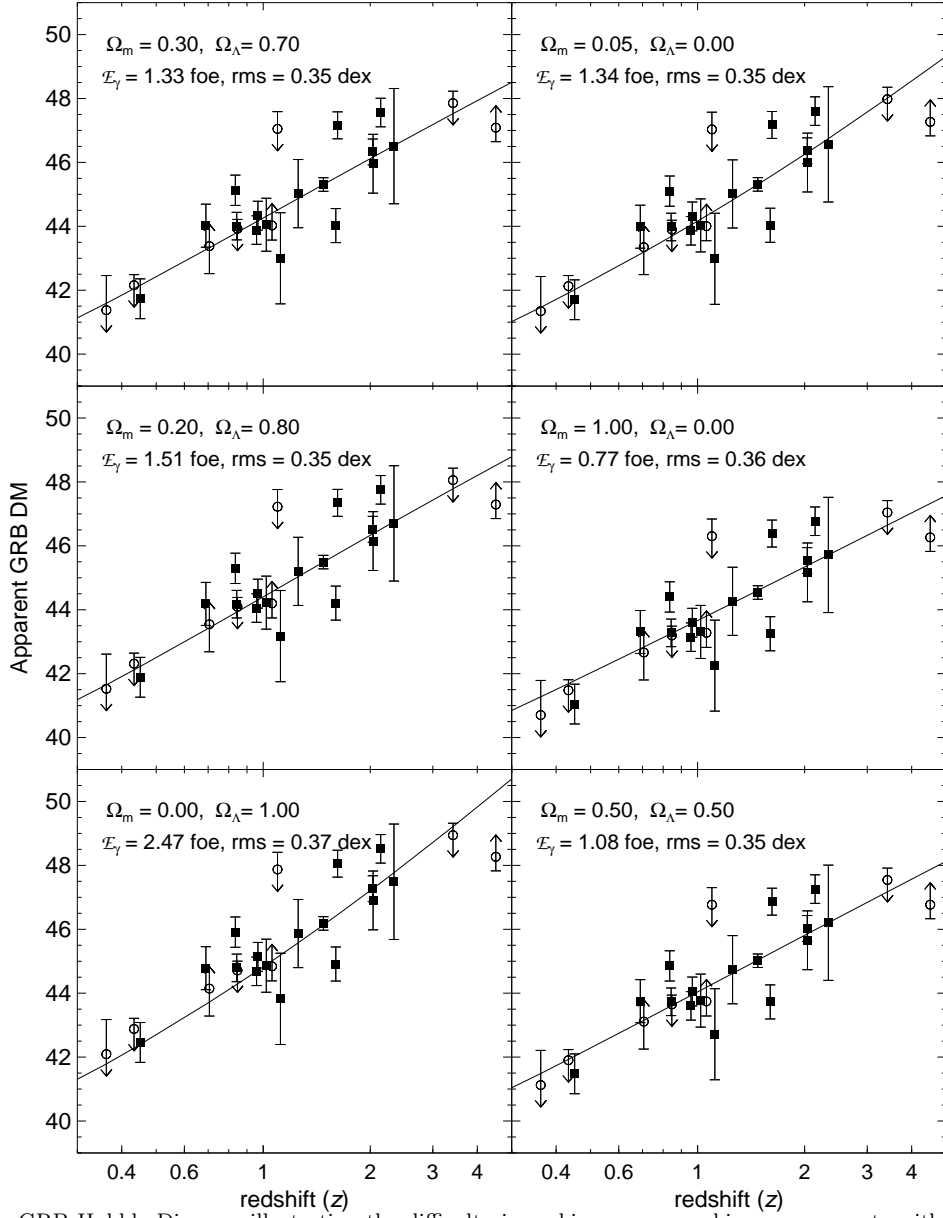,width=5.3in}}
\vskip -0.6cm \caption[]{\footnotesize The GRB Hubble Diagram illustrating the difficulty in 
making cosmographic measurements with GRBs.  Shown are Hubble Diagrams
for six different cosmologies for the 24 GRBs with known redshift and
a measurement or constraint on $t_{\rm jet}$. The solid squares
represent the apparent GRB distance modulus in the given cosmology
with associated 1\,$\sigma$ errors. Open circles show those sources
with upper or lower limits on the DM measurement. The solid curves are
the theoretical DM for the given cosmology. Though the values of
${\cal E}_\gamma$ vary by more than a factor of 3 over the cosmologies
shown, the observed variance about the theoretical curves is almost
the same ($\approx 0.35$ dex); we discuss in the text why this is so.
Clearly, the current sample cannot distinguish between a wide range of
cosmologies.  Here, and throughout the paper, we use $H_0 = 65$ km
s$^{-1}$ Mpc$^{-1}$.}
\label{fig:hub}
\end{figure*}

A Hubble diagram, apparent distance modulus versus redshift, is one
way to visualize the standard candle relation, now where the distance
in the vertical axis corresponds to magnitude (i.e.,~2.5 mag is a
difference of a factor of 10). In Figure \ref{fig:hub}, we present the
apparent GRB distance moduli for the 24 GRBs where such measurements
(or constraints thereon) are now possible. The solid curves show the
theoretical distance moduli in six cosmologies. For each cosmology, we
re-compute ${\cal E}_\gamma$ using the median of the same well-studied
bursts as described above. This value of ${\cal E}_\gamma$ is then
used to set the zeropoint of the apparent GRB DM measurements from
equation \ref{eq:dm1}.

Aside from an overall zeropoint, the values of the apparent GRB DM are
almost\footnotemark\footnotetext{
    \label{foot:cosmo}There is a small dependence of $\theta_{\rm
    jet}$ upon cosmology, such that for small opening angles, the
    value of DM goes as $1.25 \times \log_{10} rD_l(z)$, for a ratio,
    $rD_l(z)$, of luminosity distances versus redshift relation in two
    different cosmologies. Even at the extrema, for most plausible
    cosmologies, this ratio is less than about 2 for redshifts less
    than five. Therefore, at most, the DM constructed from observable
    parameters could be systematically off by $\ale 0.35$; that is,
    the role of cosmology in determining the RHS of equation 4 is
    usually small compared with the error on the observables
    themselves.}
entirely dependent on observables, and not cosmology, yet the
theoretical DM curve are sensitive to cosmology ($\propto 5 \log
d_l$). That the theoretical DM curves trace the apparent GRB DM
measurements in Figure \ref{fig:hub}, is simply a recasting of the
standard candle energy result.

\subsection{Limitations in Cosmographic Applications}

One striking feature of Figure \ref{fig:hub} is that the data appear
to be well-represented by the theoretical curves in all six (rather
different) cosmologies, with no apparent discrepant trend with
redshift. The rms scatter about the respective ${\cal E}_\gamma$ is
about the same in each cosmology ($\approx 0.35$ dex), dominated by
the outliers in the distribution. The main difference between various
cosmological formulations is that the standard energy varies by a
factor of $\sim 3$, all about energies of $\sim$1 foe.

We explain the insensitivity of the energy data to cosmographic
discrimination as follows. Without an {\em a priori} knowledge of the
true standard energy release -- either via observations of a local
sample or from theoretical studies --- the data themselves are used to
find the standard energy, setting the zeropoint for the DM
measurements. This is akin to side-stepping the cosmological distance
``ladder.''  Indeed without an anchor, the divergence of theoretical
DMs in different cosmologies at large redshifts is subtle, and hence
more difficult to observe. In other words, unlike with other
cosmological tools such as with SNe Type Ia (with local calibrations)
or the Sunyaev-Zeldovich effect (which depend only on local physics),
any cosmographic discrimination with GRBs is performed
differentially. As can be seen in Figure \ref{fig:hub}, the current
scatter in GRB energies is simply to large for any apparent
differential trend with redshift to be seen. The detection of
lower-redshift ($z \ale 0.3$) bursts with {\it Swift} would help pin
down ${\cal E}_\gamma$ and increase the sensitivity of the sample to
cosmology.

\section{Toward a More Homogeneous Set}
\label{sec:homogen}

As noted in \S \ref{sec:introduction}, cosmographic applications
benefit enormously if the objects are well understood physically.
Lacking a clear understanding (as is arguably still the case with Type
Ia SNe), an astronomical appreciation of the diversity of the
phenomenon is critical.  In short, if the scatter on $E_\gamma$ is
ever to be reduced {\em a priori}, it is essential that we develop the
ability to identify sub-classes from observations so that a
homogeneous set of GRBs can be identified for cosmographic
applications.
 
\begin{figure*}[hbt]
\centerline{\psfig{file=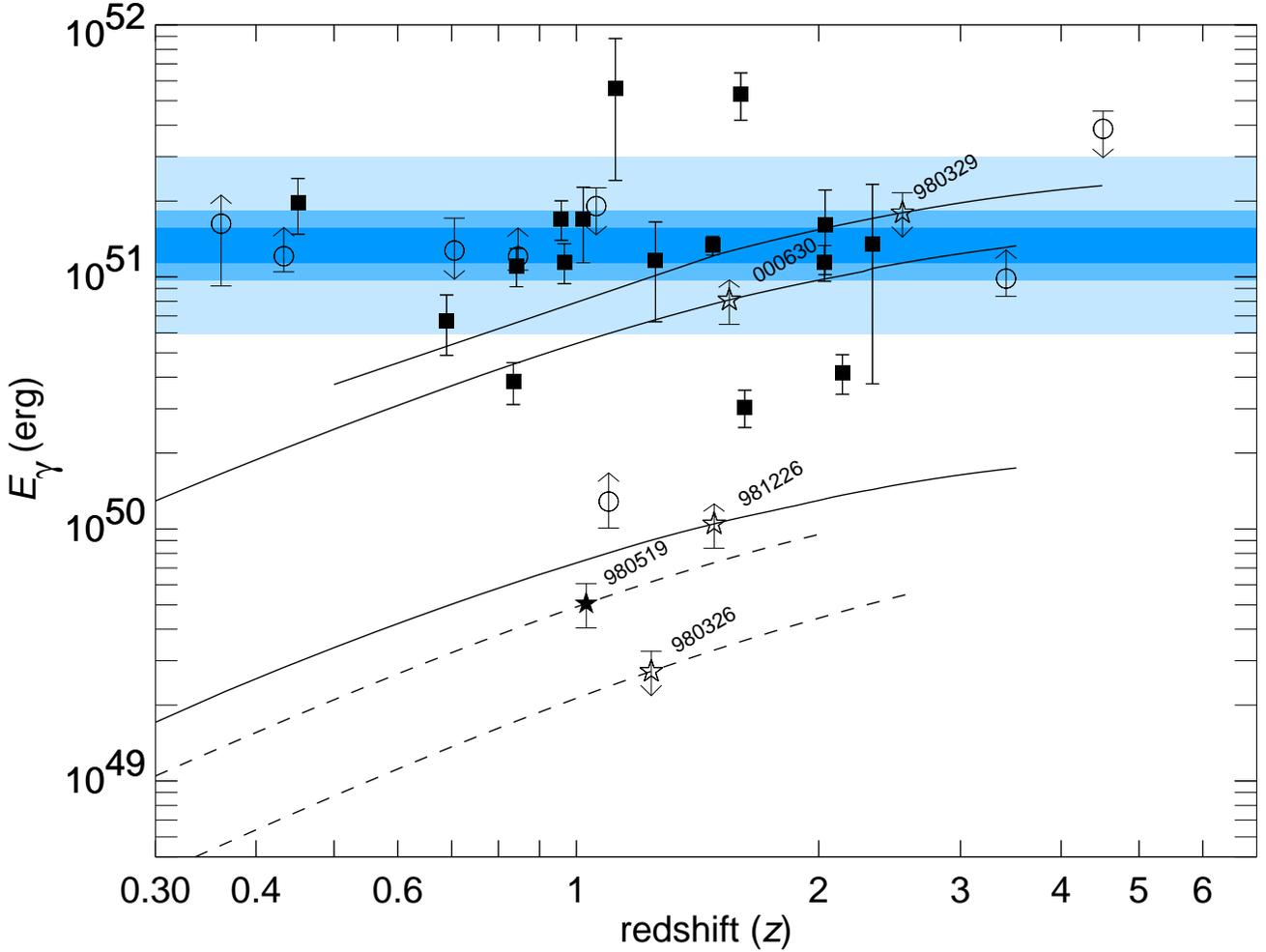,height=5.5in}}
\caption[]{GRB energy release versus redshift assuming a top-hat jet illustrating the diversity in
GRB energies. Bands of 1, 2, and 5 $\sigma$ about the mean energy
${\cal E}_\gamma = 1.33$ foe are shown. Solid box and open circle
symbols are the same as in Figure \ref{fig:hub}.  Plotted are the
trajectories of five GRBs with no known spectroscopic redshift
(labeled with star symbols). Redshift limits are place from
photometric observations of the afterglow (980519: \citealt{jhb+01};
980326: \citealt{bkd+99}; 000630: \citealt{fjg+01}; 991216:
\citealt{hum+00})  assuming no extinction from the redshifted Lyman
$\alpha$ forest in the most blue filter. The photometric redshift of
the host galaxy of 980329 is from \citet{jah+03}. While the energies
of GRB\,980329 and GRB\,000630 could be consistent with the standard
value at redshifts beyond $z \sim 1.5$, at no redshift could the
energies of GRB\,980326 and GRB\,980519 (dashed lines) be consistent
unless the densities were significantly higher than the canonical
value of 10 cm$^{-3}$. Interestingly, these two bursts are associated
with having occurred in a wind-blown density environment. While most
long-duration GRBs appear to fall within a narrow range of energies,
this shows there are several exceptional outliers, with demonstrably
different afterglow properties from other bursts; we refer to such
bursts as ``fast-fading GRBs'' or f-GRBs.}
\label{fig:hub2}
\end{figure*}

We already know of at least one member of a distinct sub-class of
GRBs: the GRB associated with SN\,1998bw \citep{gvv+98,kfw+98}. Indeed
GRB\,980425, with an isotropic-equivalent energy release of
$E_\gamma\sim 7\times 10^{47}\,$erg, is clearly abnormal (see
Figure~\ref{fig:hub2}). \citet{bkf+98} suggested that such
low-luminosity GRBs that are associated with supernovae to be
``S-GRBs''. Recently, \citet{norr02} identified $\sim$90 such
low-redshift S-GRB candidates in the {\it BATSE} sample.  Without
appealing to $\gamma$-ray properties, such GRBs are readily identified
by their low redshift, and possibly, the dominance of the associated
supernova emission over the afterglow light.

Of the more classical ``cosmological GRBs,'' those with known
redshifts beyond $\age 100$ Mpc or associated with distant galaxies,
we can use Figure 3 help identify classes of outliers.  In Figure
\ref{fig:hub2}, we show the $E_\gamma$ distribution energy versus
redshift for the 24 bursts with known redshift. We also show the
trajectory of $E_\gamma$ for the 5 bursts where no redshift is
known. These curves were found by computing the individual
$k$-corrections and $f_b$ as a function of redshift. Note that the
implied $E_\gamma$ curve begins to flatten at high redshift as a
result of the mitigation between larger $E_{\rm iso}(\gamma)$ and
smaller inferred $f_b$. From Figure \ref{fig:hub2}, it can be seen
that the energy release in GRB\,980329 and GRB\,000630 could have been
consistent with ${\cal E}_\gamma$ for redshift $z \age 1.5$. The
present observations provide only a weak lower-limit to the energy of
GRB\,981226.

\subsection{``fast-fading GRBs'' (f-GRBs)---A New Sub-Classification?}

As is readily seen in Figure \ref{fig:hub2}, GRB\,980519 and
GRB\,980326, stand out for being significantly lower in energy release
than the cosmological set.  Even at a redshift of 5, these bursts
released less than 0.2 foe. Interestingly, independent of the inferred
low energy, GRB\,980519 and GRB\,980326 have previously been
recognized as peculiar in that both afterglows exhibited an unusually
steep decline, flux $\propto t^{-\alpha}$ with $\alpha\sim 2$ (instead
of the usual $\alpha \sim 1$) at early times.  We propose that bursts
with early-time steeply declining afterglows may be intrinsically
under-energetic and suggest a nomenclature, ``fast-fading GRBs'' or
f-GRBs (in analogy with the ``Branch peculiar'' of Type Ia SNe), for
such GRBs.

While the identification of f-GRBs, based upon fast-declining
afterglows, is also an appeal to the physical state of those systems,
under the assumption of an ISM density, we cannot identify any
unifying observation of the 5 closer energy outliers (970508, 990123,
990510, 000418, 011211) that belie some unifying physical process. We
therefore cannot (yet) distinguish such bursts {\em a priori} from the
remainder of the set. 

Bursts that are f-GRBs could be either genuinely low-energy events or
GRBs which typical energy releases by taking place in particularly
dense media \citepeg{dl01}; note that for the best studied
f-GRB---030329 (see below)---there is no indication of a dense
medium. In formulating our energy measurements we have explicitly
assumed that the afterglow emission before the jet break arises in a
medium where density is constant and homogeneous (i.e., $\rho(r)
\propto r^{0}$). Since the environments of a least some GRBs are
thought to be clumpy
\citep{lrcg02,bsw+03,hp03} and wind-blown (i.e., $\rho(r) \propto
r^{-2}$; \citealt{cl99,pbr+02}), this is clearly not the case for all
GRBs in our sample; therefore, our estimations of $f_b$ may be
inappropriate for some GRBs. As in the homogeneous case, GRBs the
dynamics of $\Gamma$ in jetted GRB within winds should be insensitive
to the jet edge so long as $\Gamma \ale 1/\theta_{\rm jet}$. We can
therefore calculate $\theta_{\rm jet}$(wind) by inverting the equation
for $\Gamma$ in eq.~1 of \citet{cl00a}:
\begin{equation}
\theta_{\rm jet}({\rm wind}) = 0.169 \left(\frac{1+z}{2}\right)^{-1/4}
	\left(\frac{E_{\rm iso}(\gamma)}{10^{52}~{\rm erg}}\right)^{-1/4}
	A*^{1/4} \left(\frac{t_{\rm jet}}{{\rm day}}\right)^{1/4}.
\label{eq:wind}
\end{equation}
For all bursts in our sample for which $A*$ has been reported we
calculate $\theta_{\rm jet}({\rm wind})$ following
eqs.~\ref{eq:egamma} and \ref{eq:wind}. The results are as follows
($\theta_{\rm jet}({\rm wind})$, $E_\gamma$/$10^{50}$ erg, $A*$
reference):
970508 (20.81$^\circ$, 3.560; \citealt{pk02}); 980519 (4.88$^\circ$,
6.262; \citealt{pk02}) (5.13$^\circ$, 6.941; \citealt{cl99}); 991208
($<$5.29$^\circ$, $<$4.771; \citealt{lc01}); 990123 (2.82$^\circ$,
17.390; \citealt{pk02}); 000301C (8.15$^\circ$, 4.411;
\citealt{lc01}); 000418 (11.78$^\circ$, 15.834; \citealt{pk02});
000926 (5.24$^\circ$, 11.689; \citealt{pk02}); 010222 (1.93$^\circ$,
4.891; \citealt{pk02}); 011121 ($>$4.47$^\circ$, $>$1.390;
\citealt{pbr+02}).
(We assume $z=2.5$ for GRB\,980519. Note that the wind hypothesis for
GRB\,991208 is strongly disfavored with late-time radio modeling of
\citealt{gfs+03}.)  Comparison with Table 2 shows that the inferred
values for $\theta_{\rm jet}$ and $E_\gamma$ are usually smaller under
the wind assumption. Uncertainties in $A*$ are not generally provided
so it is unclear if the energy determinations are consistent between
winds and ISM models.  GRB\,970508 is underluminous by a factor of
$\sim$3 in both the ISM and wind cases. Both GRB\,000418 and
GRB\,990123, overluminous in the ISM case, are more consistent with
the standard ${\cal E}_\gamma$ in the wind case.

Only GRB\,980519 and GRB\,011121 show convincing evidence for a
wind-stratified medium under the assumption of a top-hat jet model
(GRB\,980326 is consistent with a wind; \citealt{bkd+99}). As seen,
the energy release in GRB\,011121 is similar in both the wind and
ISM cases. Interestingly, the energy release in GRB\,980519, again
assuming $z=2.5$ and a wind-blown circumburst medium, is 6.3 $\times
10^{50}$ erg and thus consistent with ${\cal E}_\gamma$. Therefore it
may be reasonable to suspect that all f-GRBs are bursts that occur in
wind-stratified media (but not necessarily the converse).

\section{Discussion and Conclusions}

With nine new bursts suitable for $E_\gamma$ measurement after FKS,
the evidence for a standard release of energy in GRBs persists.  By
incorporating more realistic measurements of ambient density and under
the assumption of a top-hat jet and constant ambient density media, we
find a standard candle energy release of ${\cal E}_\gamma =
1.33\,h_{65}^{-2}$ foe, with a burst--to--burst scatter of 0.35 dex
about this value. Though this result makes it interesting to consider
cosmographic applications of GRBs, we have shown that without a local
calibration of the true energy release, the observed scatter is simply
too large to make any meaningful impact on cosmographic measurements.

The most obvious way to reduce the scatter on ${\cal E}_\gamma$ is to
simply remove outliers from the sample, based purely upon discrepant
energy measurements. However, this is a dangerously circular approach
since it is particularly these apparent outliers in energy, at high
redshift especially, that place the most meaningful constraints for
cosmography.

Still, at this stage, we are confined to first recognize heterogeneity
in the context of a cosmological model and then find similarities
between discrepant bursts, based upon observations or theory that are
{\it complementary} to energy. To this end, we have noted that both
GRB\,980519 and GRB\,980326 are low-energy bursts {\it and} stand out
based upon afterglow observations.

Indeed, lacking firm theoretical predictions for sub-classification,
we have proposed that there are at least two sub-classes -- S-GRBs and
f-GRBs -- of GRBs which can be typed or classified from simple
observations.  Both groups were first identified by low inferred
$\gamma$-ray energy releases.  Now, S-GRBs are characterized by low
redshift\footnotemark\footnotetext{The origin of $\gamma$-ray emission
in S-GRBs has now been discussed in some detail
\citepeg{tmm01} and it is reasonable to infer that the $\gamma$-ray
emission arises from entirely different process as compared to the
internal shocks of cosmologically located GRBs.} and f-GRBs are
identified by the rapid fading, $t^{-2}$, of their afterglow at early
times, $t\lsim$ 0.5 day (restframe).

The advantage of empirical distance indicators (e.g.,
\citealt{sch03}) is that they can be entirely based on the $\gamma$-ray
observations and do not require (once calibrated) additional
time-consuming observations from the ground. However, as with many
empirical indicators, the method relies on poorly understood physical
mechanisms (although, see, \citealt{sal00}). The sheer bewildering
diversity in any high-energy property of GRBs -- richness of profiles,
overall profile shapes, fluences, peak luminosities and even peak
energies -- should give pause to claims of the predictive power of
distance indicators in the absence of a realistic physical model for
the emission mechanisms.

The physical uncertainty aside, it is clear that the high energy
properties can and have provided gross distinctions. The well-known
separation in energy and spectra between short-bursts and hard-bursts
\citep{kmf+93}, for example, has led to the belief that bursts from
each class may originate from different progenitors. As a more recent
example, we note that \citet{norr02} has used the lag--luminosity
relation to identify a subsample of long-duration bursts whose
properties distinguish them from the general population.  However, we
question whether empirical relations have demonstrated that they
possess the precision necessary to advance cosmography.  There is
little reason based on the limited understanding of the GRB phenomenon
to strongly justify an optimistic expectation in this direction.

The approach we have used in FKS and herein has the advantage that it
is based upon energy release, a fundamental quantity in explosions; on
physical grounds, it is reasonable to expect that such a quantity
could be standard (or at least amenable to calibration) despite the
rich diversity in GRB properties. The standard energy result is
corroborated by other studies (which do not rely on the $\gamma$-ray
data), namely multi-wavelength afterglow modeling and X-ray
observations \citep{pk01,pkpp01,bkf03}.

We stress that if GRBs are ever to be used for cosmographic purposes,
there must be a significantly increased understanding in the diversity
of the phenomenon (let alone the secular evolution with redshift). To
this end, using simple observations, at least two sub-classes of GRBs,
the S-GRBs and the f-GRBs, may be readily recognized. For the short
term, we believe that the identification of such bursts will provide
informative direction to hone in on GRB progenitors.  To weed out such
bursts and create a more homogeneous sample for cosmography, however,
our proposed identification methods have the disadvantage of requiring
significant follow up observations. Nevertheless, it is possible that
in the {\it SWIFT} era, dedicated facilities will undertake the
necessary follow-up observations quite routinely.

After this paper was submitted, GRB\,030329 was discovered and found
to fade rapidly at early times ($F_\nu \propto t^{1.97 \pm 0.12}$)
\citep{pfk+03} suggesting that the burst fit our definition of an f-GRB. 
Using the same formalism above, we find $E_\gamma = (8.18 \pm 1.00)
\times 10^{49}$ erg assuming a constant density ISM and $n=10$
cm$^{-3}$ or $1.3 \times 10^{50}$ erg assuming a wind with $A* =
1.0$. In the former case, GRB\,030329 is underluminous by a factor of
16.3 from the ${\cal E}_\gamma$ found herein. Thus GRB\,030329 joins
GRB\,980519 and GRB\,980326 in the same sub-class both on energetics
and phenomenological grounds. 

Given new the connection of GRB\,030329 with a supernova
\citep{smg+03}, we suggest that all f-GRBs may be associated with
supernovae. Obviously with afterglows that fade fast there is an
increasing likelihood of detecting a supernova component (the
non-detection of a SN component in the optical light curve of
GRB\,980519, see \citealt{jhb+01}, may simply be due to $z \age 1$ of
this burst). However, we suggest that the connection may be more than
due to observational bias; instead long-duration GRBs could arise from
two populations of bursts
\citep{cl99}, with f-GRBs from massive stars that produce winds prior
to the GRB and a concurrent supernova with the GRB
event. Alternatively, especially in light of the detection of a
supernova component with GRB\,011121 \citep{bkp+02,gsw+03}, we favor
the hypothesis that all long-duration GRBs also lead to supernovae,
with the diversity in GRB energies set by a progenitor-dependent ratio
of the energy channeled into the relativistic jets to the energy
powering the explosion the massive star progenitor.

\acknowledgments

Many thanks to the referee for illuminating comments. JSB gratefully
acknowledges a fellowship from the Harvard Society of Fellows and the
generous research support from the Harvard-Smithsonian Center for
Astrophysics. JSB also thanks Prof.~J.~M.~Paredes and the Astronomy
department at the Universitat de Barcelona for their hospitality and
funding support. SRK acknowledges support from NASA and the NSF.  We
thank A.~Friedman and R.~Perna for helpful discussions. The National
Radio Astronomy Observatory is a facility of the National Science
Foundation operated under cooperative agreement by Associated
Universities, Inc.


\begin{thebibliography}{116}
\expandafter\ifx\csname natexlab\endcsname\relax\def\natexlab#1{#1}\fi

\bibitem[{{Amati} {et~al.}(2000){Amati}, {Frontera}, {Vietri}, {in't Zand},
  {Soffitta}, {Costa}, {Del Sordo}, {Pian}, {Piro}, {Antonelli}, {Fiume},
  {Feroci}, {Gandolfi}, {Guidorzi}, {Heise}, {Kuulkers}, {Masetti},
  {Montanari}, {Nicastro}, {Orlandini}, \& {Palazzi}}]{afv+00}
{Amati}, L. {\it et al.} 2000, Science, 290, 953

\bibitem[{{Andersen} {et~al.}(2000){Andersen}, {Hjorth}, {Pedersen}, {Jensen},
  {Hunt}, {Gorosabel}, {M{\o}ller}, {Fynbo}, {Kippen}, {Thomsen}, {Olsen},
  {Christensen}, {Vestergaard}, {Masetti}, {Palazzi}, {Hurley}, {Cline},
  {Kaper}, \& {Jaunsen}}]{ahp+00}
{Andersen}, M.~I. {\it et al.} 2000, A\&A, 364, L54

\bibitem[{{Barth} {et~al.}(2003){Barth}, {Sari}, {Cohen}, {Goodrich}, {Price},
  {Fox}, {Bloom}, {Soderberg}, \& {Kulkarni}}]{bsc+03}
{Barth}, A.~J. {\it et al.} 2003, ApJ ({\it Letters}), 584, L47

\bibitem[{{Berger} {et~al.}(2001){Berger}, {Diercks}, {Frail}, {Kulkarni},
  {Bloom}, {Sari}, {Halpern}, {Mirabal}, {Taylor}, {Hurley}, {Pooley},
  {Becker}, {Wagner}, {Terndrup}, {Statler}, {Wik}, {Mazets}, \&
  {Cline}}]{bdf+01}
{Berger}, E. {\it et al.} 2001, ApJ, 556, 556

\bibitem[{Berger {et~al.}(2003)Berger, Kulkarni, \& Frail}]{bkf03}
Berger, E., Kulkarni, S.~R., and Frail, D.~A. 2003, in preparation

\bibitem[{{Berger} {et~al.}(2000){Berger}, {Sari}, {Frail}, {Kulkarni},
  {Bertoldi}, {Peck}, {Menten}, {Shepherd}, {Moriarty-Schieven}, {Pooley},
  {Bloom}, {Diercks}, {Galama}, \& {Hurley}}]{bsf+00}
{Berger}, E. {\it et al.} 2000, ApJ, 545, 56

\bibitem[{Bersier {et~al.}(2003)Bersier, Stanek, Winn, Grav, Holman, Matheson,
  Mochejska, Steeghs, Walker, Garnavich, Quinn, Jha, Calitz, \&
  Meintjes}]{bsw+03}
Bersier, D. {\it et al.} 2003, {astro-ph/0211130}

\bibitem[{Blandford \& McKee(1976)}]{bm76}
Blandford, R.~D. and McKee, C.~F. 1976, Phys. of Fluids, 19, 1130

\bibitem[{Bloom {et~al.}(2003)Bloom, Berger, Kulkarni, Djorgovski, \&
  Frail}]{bbk+03}
Bloom, J.~S., Berger, E., Kulkarni, S.~R., Djorgovski, S.~G., and Frail, D.~A.
  2003, {AJ}, in press; astro-ph/0212123

\bibitem[{{Bloom} {et~al.}(2001{\natexlab{a}}){Bloom}, {Djorgovski}, \&
  {Kulkarni}}]{bdk01}
{Bloom}, J.~S., {Djorgovski}, S.~G., and {Kulkarni}, S.~R. 2001{\natexlab{a}},
  ApJ, 554, 678

\bibitem[{Bloom {et~al.}(1998)Bloom, Djorgovski, Kulkarni, \& Frail}]{bdkf98}
Bloom, J.~S., Djorgovski, S.~G., Kulkarni, S.~R., and Frail, D.~A. 1998, ApJ
  ({\it Letters}), 507, L25

\bibitem[{{Bloom} {et~al.}(2001{\natexlab{b}}){Bloom}, {Frail}, \&
  {Sari}}]{bfs01}
{Bloom}, J.~S., {Frail}, D.~A., and {Sari}, R. 2001{\natexlab{b}}, AJ, 121,
  2879

\bibitem[{{Bloom} {et~al.}(1999){Bloom}, {Kulkarni}, {Djorgovski},
  {Eichelberger}, {Cote}, {Blakeslee}, {Odewahn}, {Harrison}, {Frail},
  {Filippenko}, {Leonard}, {Riess}, {Spinrad}, {Stern}, {Bunker}, {Dey},
  {Grossan}, {Perlmutter}, {Knop}, {Hook}, \& {Feroci}}]{bkd+99}
{Bloom}, J.~S. {\it et al.} 1999, Nature, 401, 453

\bibitem[{{Bloom} {et~al.}(1998){Bloom}, {Kulkarni}, {Harrison}, {Prince},
  {Phinney}, \& {Frail}}]{bkf+98}
{Bloom}, J.~S., {Kulkarni}, S.~R., {Harrison}, F., {Prince}, T., {Phinney},
  E.~S., and {Frail}, D.~A. 1998, ApJ ({\it Letters}), 506, L105

\bibitem[{{Bloom} {et~al.}(2002){Bloom}, {Kulkarni}, {Price}, {Reichart},
  {Galama}, {Schmidt}, {Frail}, {Berger}, {McCarthy}, {Chevalier}, {Wheeler},
  {Halpern}, {Fox}, {Djorgovski}, {Harrison}, {Sari}, {Axelrod}, {Kimble},
  {Holtzman}, {Hurley}, {Frontera}, {Piro}, \& {Costa}}]{bkp+02}
{Bloom}, J.~S. {\it et al.} 2002, ApJ ({\it Letters}), 572, L45

\bibitem[{{Branch} {et~al.}(1993){Branch}, {Fisher}, \& {Nugent}}]{bfn93}
{Branch}, D., {Fisher}, A., and {Nugent}, P. 1993, AJ, 106, 2383

\bibitem[{Castro {et~al.}(2000{\natexlab{a}})}]{cdk+00}
Castro, M. {\it et al.} 2000{\natexlab{a}}, {{GCN} notice 851}

\bibitem[{Castro {et~al.}(2000{\natexlab{b}})Castro, Diercks, Djorgovski,
  Kulkarni, Galama, Bloom, Harrison, \& Frail}]{cdd+00}
Castro, S.~M., Diercks, A., Djorgovski, S.~G., Kulkarni, S.~R., Galama, T.~J.,
  Bloom, J.~S., Harrison, F.~A., and Frail, D.~A. 2000{\natexlab{b}}, {GCN}
  notice 605

\bibitem[{{Castro-Tirado} {et~al.}(2001){Castro-Tirado}, {Sokolov},
  {Gorosabel}, {Castro Cer{\' o}n}, {Greiner}, {Wijers}, {Jensen}, {Hjorth},
  {Toft}, {Pedersen}, {Palazzi}, {Pian}, {Masetti}, {Sagar}, {Mohan}, {Pandey},
  {Pandey}, {Dodonov}, {Fatkhullin}, {Afanasiev}, {Komarova}, {Moiseev},
  {Hudec}, {Simon}, {Vreeswijk}, {Rol}, {Klose}, {Stecklum}, {Zapatero-Osorio},
  {Caon}, {Blake}, {Wall}, {Heinlein}, {Henden}, {Benetti}, {Magazz{\` u}},
  {Ghinassi}, {Tommasi}, {Bremer}, {Kouveliotou}, {Guziy}, {Shlyapnikov},
  {Hopp}, {Feulner}, {Dreizler}, {Hartmann}, {Boehnhardt}, {Paredes}, {Mart{\'
  i}}, {Xanthopoulos}, {Kristen}, {Smoker}, \& {Hurley}}]{csg+01}
{Castro-Tirado}, A.~J. {\it et al.} 2001, A\&A, 370, 398

\bibitem[{{Chevalier} \& {Li}(2000)}]{cl00a}
{Chevalier}, R.~A. and {Li}, Z. 2000, ApJ, 536, 195

\bibitem[{{Chevalier} \& {Li}(1999)}]{cl99}
{Chevalier}, R.~A. and {Li}, Z.-Y. 1999, ApJ ({\it Letters}), 520, L29

\bibitem[{Crew {et~al.}(2002)}]{cvv+02}
Crew {\it et al.} 2002, {GCN} notice 1734

\bibitem[{{Dai} \& {Lu}(2001)}]{dl01}
{Dai}, Z.~G. and {Lu}, T. 2001, A\&A, 367, 501

\bibitem[{Djorgovski {et~al.}(2000)Djorgovski, Bloom, \& Kulkarni}]{dbk00}
Djorgovski, S.~G., Bloom, J.~S., and Kulkarni, S.~R. 2000, {ApJ Lett.},
  accepted; astro-ph/0008029

\bibitem[{{Djorgovski} {et~al.}(2001){Djorgovski}, {Frail}, {Kulkarni},
  {Bloom}, {Odewahn}, \& {Diercks}}]{dfk+01}
{Djorgovski}, S.~G., {Frail}, D.~A., {Kulkarni}, S.~R., {Bloom}, J.~S.,
  {Odewahn}, S.~C., and {Diercks}, A. 2001, ApJ, 562, 654

\bibitem[{{Djorgovski} {et~al.}(1998){Djorgovski}, {Kulkarni}, {Bloom},
  {Goodrich}, {Frail}, {Piro}, \& {Palazzi}}]{dkb+98b}
{Djorgovski}, S.~G., {Kulkarni}, S.~R., {Bloom}, J.~S., {Goodrich}, R.,
  {Frail}, D.~A., {Piro}, L., and {Palazzi}, E. 1998, ApJ ({\it Letters}), 508,
  L17

\bibitem[{Fenimore \& Ramirez-Ruiz(2003)}]{frr03}
Fenimore, E.~E. and Ramirez-Ruiz, E. 2003, submitted; astro-ph/0004176

\bibitem[{Fox {et~al.}(2002)}]{fox+02}
Fox {\it et al.} 2002, in preparation

\bibitem[{Frail {et~al.}(2003)}]{fyb+03}
Frail {\it et al.} 2003, in preparation

\bibitem[{{Frail} {et~al.}(1999){Frail}, {Kulkarni}, {Bloom}, {Djorgovski},
  {Gorjian}, {Gal}, {Meltzer}, {Sari}, {Chaffee}, {Goodrich}, {Frontera}, \&
  {Costa}}]{fkb+99}
{Frail}, D.~A. {\it et al.} 1999, ApJ ({\it Letters}), 525, L81

\bibitem[{{Frail} {et~al.}(2001){Frail}, {Kulkarni}, {Sari}, {Djorgovski},
  {Bloom}, {Galama}, {Reichart}, {Berger}, {Harrison}, {Price}, {Yost},
  {Diercks}, {Goodrich}, \& {Chaffee}}]{fks+01}
---. 2001, ApJ ({\it Letters}), 562, L55

\bibitem[{{Frail} {et~al.}(2000{\natexlab{a}}){Frail}, {Kulkarni}, {Sari},
  {Taylor}, {Shepherd}, {Bloom}, {Young}, {Nicastro}, \& {Masetti}}]{fks+00}
---. 2000{\natexlab{a}}, ApJ, 534, 559

\bibitem[{{Frail} {et~al.}(2000{\natexlab{b}}){Frail}, {Waxman}, \&
  {Kulkarni}}]{fwk00}
{Frail}, D.~A., {Waxman}, E., and {Kulkarni}, S.~R. 2000{\natexlab{b}}, ApJ,
  537, 191

\bibitem[{{Frontera} {et~al.}(1998){Frontera}, {Amati}, {Costa}, {Feroci},
  {Muller}, {Pizzichini}, {Cinti}, {dal Fiume}, {Heise}, {Nicastro},
  {Orlandini}, {Palazzi}, \& {in 't Zand}}]{fac+98}
{Frontera}, F. {\it et al.} 1998, in Gamma-Ray Bursts, 446

\bibitem[{{Frontera} {et~al.}(2000){Frontera}, {Antonelli}, {Amati},
  {Montanari}, {Costa}, {Dal Fiume}, {Giommi}, {Feroci}, {Gennaro}, {Heise},
  {Masetti}, {Muller}, {Nicastro}, {Orlandini}, {Palazzi}, {Pian}, {Piro},
  {Soffitta}, {Stornelli}, {in 't Zand}, {Frail}, {Kulkarni}, \&
  {Vietri}}]{faa+00}
{Frontera}, F. {\it et al.} 2000, ApJ, 540, 697

\bibitem[{Frontera {et~al.}(2000)}]{fro00}
Frontera, F. {\it et al.} 2000, {ApJ} (Lett.), submitted

\bibitem[{Frontera {et~al.}(2001)}]{fag+02}
---. 2001, {GCN} notice 1215

\bibitem[{Fruchter {et~al.}(2000)Fruchter, Vreeswijk, Hook, \& Pian}]{fvhp00}
Fruchter, A., Vreeswijk, P., Hook, R., and Pian, E. 2000, {GCN} notice 752

\bibitem[{{Fynbo} {et~al.}(2001){Fynbo}, {Jensen}, {Gorosabel}, {Hjorth},
  {Pedersen}, {M{\o}ller}, {Abbott}, {Castro-Tirado}, {Delgado}, {Greiner},
  {Henden}, {Magazz{\` u}}, {Masetti}, {Merlino}, {Masegosa}, {{\O}stensen},
  {Palazzi}, {Pian}, {Schwarz}, {Cline}, {Guidorzi}, {Goldsten}, {Hurley},
  {Mazets}, {McClanahan}, {Montanari}, {Starr}, \& {Trombka}}]{fjg+01}
{Fynbo}, J.~P.~U. {\it et al.} 2001, A\&A, 369, 373

\bibitem[{{Galama} {et~al.}(2003{\natexlab{a}}){Galama}, {Frail}, {Sari},
  {Berger}, {Taylor}, \& {Kulkarni}}]{gfs+03}
{Galama}, T.~J., {Frail}, D.~A., {Sari}, R., {Berger}, E., {Taylor}, G.~B., and
  {Kulkarni}, S.~R. 2003{\natexlab{a}}, ApJ, 585, 899

\bibitem[{{Galama} {et~al.}(2003{\natexlab{b}}){Galama}, {Reichart}, {Brown},
  {Kimble}, {Price}, {Berger}, {Frail}, {Kulkarni}, {Yost}, {Gal-Yam}, {Bloom},
  {Harrison}, {Sari}, {Fox}, \& {Djorgovski}}]{grb+03}
{Galama}, T.~J. {\it et al.} 2003{\natexlab{b}}, ApJ, 587, 135

\bibitem[{{Galama} {et~al.}(1998){Galama}, {Vreeswijk}, {Van Paradijs},
  {Kouveliotou}, {Augusteijn}, {Bohnhardt}, {Brewer}, {Doublier}, {Gonzalez},
  {Leibundgut}, {Lidman}, {Hainaut}, {Patat}, {Heise}, {In 't Zand}, {Hurley},
  {Groot}, {Strom}, {Mazzali}, {Iwamoto}, {Nomoto}, {Umeda}, {Nakamura},
  {Young}, {Suzuki}, {Shigeyama}, {Koshut}, {Kippen}, {Robinson}, {De Wildt},
  {Wijers}, {Tanvir}, {Greiner}, {Pian}, {Palazzi}, {Frontera}, {Masetti},
  {Nicastro}, {Feroci}, {Costa}, {Piro}, {Peterson}, {Tinney}, {Boyle},
  {Cannon}, {Stathakis}, {Sadler}, {Begam}, \& {Ianna}}]{gvv+98}
---. 1998, Nature, 395, 670

\bibitem[{{Galama} {et~al.}(1999){Galama}, {Vreeswijk}, {van Paradijs},
  {Kouveliotou}, {Augusteijn}, {Patat}, {Heise}, {in 't Zand}, {Groot},
  {Wijers}, {Pian}, {Palazzi}, {Frontera}, \& {Masetti}}]{gvvk+99}
---. 1999, A\&A, 138, 465

\bibitem[{{Garnavich} {et~al.}(2003){Garnavich}, {Stanek}, {Wyrzykowski},
  {Infante}, {Bendek}, {Bersier}, {Holland}, {Jha}, {Matheson}, {Kirshner},
  {Krisciunas}, {Phillips}, \& {Carlberg}}]{gsw+03}
{Garnavich}, P.~M. {\it et al.} 2003, ApJ, 582, 924

\bibitem[{{Granot} \& {Sari}(2002)}]{gs02}
{Granot}, J. and {Sari}, R. 2002, ApJ, 568, 820

\bibitem[{{Groot} {et~al.}(1998){Groot}, {Galama}, {Vreeswijk}, {Wijers},
  {Pian}, {Palazzi}, {Van Paradijs}, {Kouveliotou}, {In 't Zand}, {Heise},
  {Robinson}, {Tanvir}, {Lidman}, {Tinney}, {Keane}, {Briggs}, {Hurley},
  {Gonzalez}, {Hall}, {Smith}, {Covarrubias}, {Jonker}, {Casares}, {Frontera},
  {Feroci}, {Piro}, {Costa}, {Smith}, {Jones}, {Windridge}, {Bland-Hawthorn},
  {Veilleux}, {Garcia}, {Brown}, {Stanek}, {Castro-Tirado}, {Gorosabel},
  {Greiner}, {Jaeger}, {Bohm}, \& {Fricke}}]{ggv+98c}
{Groot}, P.~J. {\it et al.} 1998, ApJ ({\it Letters}), 502, L123

\bibitem[{Halpern \& Fesen(1998)}]{hf98}
Halpern, J.~P. and Fesen, R. 1998, {GCN} notice 134

\bibitem[{{Halpern} {et~al.}(2000){Halpern}, {Uglesich}, {Mirabal}, {Kassin},
  {Thorstensen}, {Keel}, {Diercks}, {Bloom}, {Harrison}, {Mattox}, \&
  {Eracleous}}]{hum+00}
{Halpern}, J.~P. {\it et al.} 2000, ApJ, 543, 697

\bibitem[{{Harrison} {et~al.}(1999){Harrison}, {Bloom}, {Frail}, {Sari},
  {Kulkarni}, {Djorgovski}, {Axelrod}, {Mould}, {Schmidt}, {Wieringa}, {Wark},
  {Subrahmanyan}, {McConnell}, {McCarthy}, {Schaefer}, {McMahon}, {Markze},
  {Firth}, {Soffitta}, \& {Amati}}]{hbf+99}
{Harrison}, F.~A. {\it et al.} 1999, ApJ ({\it Letters}), 523, L121

\bibitem[{{Harrison} {et~al.}(2001){Harrison}, {Yost}, {Sari}, {Berger},
  {Galama}, {Holtzman}, {Axelrod}, {Bloom}, {Chevalier}, {Costa}, {Diercks},
  {Djorgovski}, {Frail}, {Frontera}, {Hurley}, {Kulkarni}, {McCarthy}, {Piro},
  {Pooley}, {Price}, {Reichart}, {Ricker}, {Shepherd}, {Schmidt}, {Walter}, \&
  {Wheeler}}]{hys+01}
---. 2001, ApJ, 559, 123

\bibitem[{Heyl \& Perna(2003)}]{hp03}
Heyl, J.~S. and Perna, R. 2003, {astro-ph/0211256}

\bibitem[{{Holland} {et~al.}(2002){Holland}, {Soszy{\' n}ski}, {Gladders},
  {Barrientos}, {Berlind}, {Bersier}, {Garnavich}, {Jha}, \& {Stanek}}]{hsg+02}
{Holland}, S.~T. {\it et al.} 2002, AJ, 124, 639

\bibitem[{{Hurley} {et~al.}(2000){Hurley}, {Cline}, {Mazets}, {Aptekar},
  {Golenetskii}, {Frederiks}, {Frail}, {Kulkarni}, {Trombka}, {McClanahan},
  {Starr}, \& {Goldsten}}]{hcm+00}
{Hurley}, K. {\it et al.} 2000, ApJ ({\it Letters}), 534, L23

\bibitem[{Hurley {et~al.}(2002)}]{hcm+02}
Hurley, K. {\it et al.} 2002, {GCN} notice 1483

\bibitem[{{in' t Zand} {et~al.}(2001){in' t Zand}, {Kuiper}, {Amati},
  {Antonelli}, {Butler}, {Costa}, {Feroci}, {Frontera}, {Gandolfi}, {Guidorzi},
  {Heise}, {Kaptein}, {Kuulkers}, {Nicastro}, {Piro}, {Soffitta}, \&
  {Tavani}}]{zka+01}
{in' t Zand}, J. . J.~M. {\it et al.} 2001, ApJ, 559, 710

\bibitem[{{Jaunsen} {et~al.}(2003){Jaunsen}, {Andersen}, {Hjorth}, {Fynbo},
  {Holland}, {Thomsen}, {Gorosabel}, {Schaefer}, {Bj{\" o}rnsson}, {Natarajan},
  \& {Tanvir}}]{jah+03}
{Jaunsen}, A.~O. {\it et al.} 2003, A\&A, 402, 125

\bibitem[{{Jaunsen} {et~al.}(2001{\natexlab{a}}){Jaunsen}, {Hjorth}, {Bj{\"
  o}rnsson}, {Andersen}, {Pedersen}, {Kjernsmo}, {Korhonen}, {S{\o}rensen}, \&
  {Palazzi}}]{jhb+01}
---. 2001{\natexlab{a}}, ApJ, 546, 127

\bibitem[{{Jaunsen} {et~al.}(2001{\natexlab{b}}){Jaunsen}, {Hjorth},
  {Bj{\"o}rnsson}, {Andersen}, {Pedersen}, {Kjernsmo}, {Korhonen},
  {S{\o}rensen}, \& {Palazzi}}]{jhb+00}
---. 2001{\natexlab{b}}, ApJ, 546, 127

\bibitem[{{Jha} {et~al.}(2001){Jha}, {Pahre}, {Garnavich}, {Calkins},
  {Kilgard}, {Matheson}, {McDowell}, {Roll}, \& {Stanek}}]{jpg+01}
{Jha}, S. {\it et al.} 2001, ApJ ({\it Letters}), 554, L155

\bibitem[{Kouveliotou {et~al.}(1993)Kouveliotou, Meegan, Fishman, Bhat, Briggs,
  Koshut, Paciesas, \& Pendleton}]{kmf+93}
Kouveliotou, C., Meegan, C.~A., Fishman, G.~J., Bhat, N.~P., Briggs, M.~S.,
  Koshut, T.~M., Paciesas, W.~S., and Pendleton, G.~N. 1993, ApJ ({\it
  Letters}), 413, 101

\bibitem[{{Kulkarni} {et~al.}(1998){Kulkarni}, {Djorgoski}, {Ramaprakash},
  {Goodrich}, {Bloom}, {Adelberger}, {Kundic}, {Lubin}, {Frail}, {Frontera},
  {Feroci}, {Nicastro}, {Barth}, {Davis}, {Filippenko}, \& {Newman}}]{kdr+98}
{Kulkarni}, S.~R. {\it et al.} 1998, Nature, 393, 35

\bibitem[{{Kulkarni} {et~al.}(1999){Kulkarni}, {Djorgovski}, {Odewahn},
  {Bloom}, {Gal}, {Koresko}, {Harrison}, {Lubin}, {Armus}, {Sari},
  {Illingworth}, {Kelson}, {Magee}, {Van Dokkum}, {Frail}, {Mulchaey},
  {Malkan}, {MCClean}, {Teplitz}, {Koerner}, {Kirkpatrick}, {Kobayashi},
  {Yadigaroglu}, {Halpern}, {Piran}, {Goodrich}, {Chaffee}, {Feroci}, \&
  {Costa}}]{kdo+99}
---. 1999, Nature, 398, 389

\bibitem[{Kulkarni {et~al.}(1998)Kulkarni, Frail, Wieringa, Ekers, Sadler,
  Wark, Higdon, Phinney, \& Bloom}]{kfw+98}
Kulkarni, S.~R. {\it et al.} 1998, Nature, 395, 663

\bibitem[{Kumar \& Granot(2003)}]{kg03}
Kumar, P. and Granot, J. 2003, {ApJ, in press.}

\bibitem[{Lamb {et~al.}(2002)}]{lra+02}
Lamb, D. {\it et al.} 2002, {GCN} notice 1600

\bibitem[{{Lazzati} {et~al.}(2002){Lazzati}, {Rossi}, {Covino}, {Ghisellini},
  \& {Malesani}}]{lrcg02}
{Lazzati}, D., {Rossi}, E., {Covino}, S., {Ghisellini}, G., and {Malesani}, D.
  2002, A\&A, 396, L5

\bibitem[{{Le Floc'h} {et~al.}(2002){Le Floc'h}, {Duc}, {Mirabel}, {Sanders},
  {Bosch}, {Rodrigues}, {Courvoisier}, {Mereghetti}, \& {Melnick}}]{fdm+02}
{Le Floc'h}, E. {\it et al.} 2002, ApJ, 581, L81

\bibitem[{{Li} {et~al.}(2001){Li}, {Filippenko}, {Treffers}, {Riess}, {Hu}, \&
  {Qiu}}]{lft+01}
{Li}, W., {Filippenko}, A.~V., {Treffers}, R.~R., {Riess}, A.~G., {Hu}, J., and
  {Qiu}, Y. 2001, ApJ, 546, 734

\bibitem[{{Li} \& {Chevalier}(2001)}]{lc01}
{Li}, Z. and {Chevalier}, R.~A. 2001, ApJ, 551, 940

\bibitem[{{Loeb} \& {Barkana}(2001)}]{lb01}
{Loeb}, A. and {Barkana}, R. 2001, Ann. Rev. Astr. Ap., 39, 19

\bibitem[{{Masetti} {et~al.}(2000){Masetti}, {Palazzi}, {Pian}, {Hunt},
  {M{\'e}ndez}, {Frontera}, {Amati}, {Vreeswijk}, {Rol}, {Galama}, {van
  Paradijs}, {Antonelli}, {Nicastro}, {Feroci}, {Marconi}, {Piro}, {Costa},
  {Kouveliotou}, {Castro-Tirado}, {Falomo}, {Augusteijn}, {B{\"o}hnhardt},
  {Lidman}, {Vanzi}, {Merrill}, {Kaminsky}, {van der Klis}, {Heemskerk}, {van
  der Hooft}, {Kuulkers}, {Pedersen}, \& {Benetti}}]{mpp+00}
{Masetti}, N. {\it et al.} 2000, A\&A, 354, 473

\bibitem[{{Matheson} {et~al.}(2002){Matheson}, {Garnavich}, {Foltz}, {West},
  {Williams}, {Falco}, {Calkins}, {Castander}, {Gawiser}, {Jha}, {Bersier}, \&
  {Stanek}}]{mgf+02}
{Matheson}, T. {\it et al.} 2002, {ApJ Letters, accepted}

\bibitem[{{M\'esz\'aros} {et~al.}(1998){M\'esz\'aros}, {Rees}, \&
  {Wijers}}]{mrw98b}
{M\'esz\'aros}, P., {Rees}, M.~J., and {Wijers}, R. A. M.~J. 1998, ApJ, 499,
  301

\bibitem[{Metzger {et~al.}(1997)Metzger, Djorgovski, Kulkarni, Steidel,
  Adelberger, Frail, Costa, \& Fronterra}]{mdk+97}
Metzger, M.~R., Djorgovski, S.~G., Kulkarni, S.~R., Steidel, C.~C., Adelberger,
  K.~L., Frail, D.~A., Costa, E., and Fronterra, F. 1997, Nature, 387, 879

\bibitem[{{Mirabal} {et~al.}(2002){Mirabal}, {Halpern}, {Kulkarni}, {Castro},
  {Bloom}, {Djorgovski}, {Galama}, {Harrison}, {Frail}, {Price}, {Reichart},
  {Ebeling}, {Bunker}, {Dawson}, {Dey}, {Spinrad}, \& {Stern}}]{mhk+02}
{Mirabal}, N. {\it et al.} 2002, ApJ, 578, 818

\bibitem[{{Norris}(2002)}]{norr02}
{Norris}, J.~P. 2002, ApJ, 579, 386

\bibitem[{{Norris} {et~al.}(2000){Norris}, {Marani}, \& {Bonnell}}]{nmb00}
{Norris}, J.~P., {Marani}, G.~F., and {Bonnell}, J.~T. 2000, ApJ, 534, 248

\bibitem[{{Panaitescu} \& {Kumar}(2001)}]{pk01}
{Panaitescu}, A. and {Kumar}, P. 2001, ApJ ({\it Letters}), 560, L49

\bibitem[{{Panaitescu} \& {Kumar}(2002)}]{pk02}
---. 2002, ApJ, 571, 779

\bibitem[{Panaitescu \& Kumar(2003)}]{pk03}
Panaitescu, A. and Kumar, P. 2003, {ApJ, in press.}

\bibitem[{{Panaitescu} {et~al.}(1998){Panaitescu}, {M\'esz\'aros}, \&
  {Rees}}]{pmr98}
{Panaitescu}, A., {M\'esz\'aros}, P., and {Rees}, M.~J. 1998, ApJ, 503, 314

\bibitem[{Pandey {et~al.}(2002)}]{psr+02}
Pandey, S.~B. {\it et al.} 2002, {Submitted} to BASI, astro-ph/0211108

\bibitem[{Piran {et~al.}(2000)Piran, Jimenez, \& Band}]{pjb00}
Piran, T., Jimenez, R., and Band, D. 2000, in Gamma Ray Bursts: 5th Huntsville
  Symposium, ed. R.~S.~M. R.~M.~Kippen \& G.~J. Fishman (Woodbury, New York:
  AIP), {87--91}

\bibitem[{{Piran} {et~al.}(2001){Piran}, {Kumar}, {Panaitescu}, \&
  {Piro}}]{pkpp01}
{Piran}, T., {Kumar}, P., {Panaitescu}, A., and {Piro}, L. 2001, ApJ ({\it
  Letters}), 560, L167

\bibitem[{{Piro} {et~al.}(2002){Piro}, {Frail}, {Gorosabel}, {Garmire},
  {Soffitta}, {Amati}, {Andersen}, {Antonelli}, {Berger}, {Frontera}, {Fynbo},
  {Gandolfi}, {Garcia}, {Hjorth}, {Zand}, {Jensen}, {Masetti}, {M{\o}ller},
  {Pedersen}, {Pian}, \& {Wieringa}}]{pfg+02}
{Piro}, L. {\it et al.} 2002, ApJ, 577, 680

\bibitem[{Price {et~al.}(2003{\natexlab{a}})}]{pfk+03}
Price, P. {\it et al.} 2003{\natexlab{a}}, {submitted to Nature}

\bibitem[{{Price} {et~al.}(2002{\natexlab{a}}){Price}, {Berger}, {Kulkarni},
  {Djorgovski}, {Fox}, {Mahabal}, {Hurley}, {Bloom}, {Frail}, {Galama},
  {Harrison}, {Morrison}, {Reichart}, {Yost}, {Sari}, {Axelrod}, {Cline},
  {Golenetskii}, {Mazets}, {Schmidt}, \& {Trombka}}]{pbk+02}
{Price}, P.~A. {\it et al.} 2002{\natexlab{a}}, ApJ, 573, 85

\bibitem[{{Price} {et~al.}(2002{\natexlab{b}}){Price}, {Berger}, {Reichart},
  {Kulkarni}, {Yost}, {Subrahmanyan}, {Wark}, {Wieringa}, {Frail}, {Bailey},
  {Boyle}, {Corbett}, {Gunn}, {Ryder}, {Seymour}, {Koviak}, {McCarthy},
  {Phillips}, {Axelrod}, {Bloom}, {Djorgovski}, {Fox}, {Galama}, {Harrison},
  {Hurley}, {Sari}, {Schmidt}, {Brown}, {Cline}, {Frontera}, {Guidorzi}, \&
  {Montanari}}]{pbr+02}
---. 2002{\natexlab{b}}, ApJ ({\it Letters}), 572, L51

\bibitem[{{Price} {et~al.}(2001){Price}, {Harrison}, {Galama}, {Reichart},
  {Axelrod}, {Berger}, {Bloom}, {Busche}, {Cline}, {Diercks}, {Djorgovski},
  {Frail}, {Gal-Yam}, {Halpern}, {Holtzman}, {Hunt}, {Hurley}, {Jacoby},
  {Kimble}, {Kulkarni}, {Mirabal}, {Morrison}, {Ofek}, {Pevunova}, {Sari},
  {Schmidt}, {Turnshek}, \& {Yost}}]{phg+01}
---. 2001, ApJ ({\it Letters}), 549, L7

\bibitem[{{Price} {et~al.}(2002{\natexlab{c}}){Price}, {Kulkarni}, {Berger},
  {Djorgovski}, {Frail}, {Mahabal}, {Fox}, {Harrison}, {Bloom}, {Yost},
  {Reichart}, {Henden}, {Ricker}, {van der Spek}, {Hurley}, {Atteia}, {Kawai},
  {Fenimore}, \& {Graziani}}]{pkb+02}
---. 2002{\natexlab{c}}, ApJ ({\it Letters}), 571, L121

\bibitem[{Price {et~al.}(2002)}]{pks+02}
Price, P.~A. {\it et al.} 2002, {ApJ (Lett)}, in press, astro-ph/0207187

\bibitem[{Price {et~al.}(2003{\natexlab{b}})}]{pkb+03}
---. 2003{\natexlab{b}}, {ApJ (Lett)} in press, astro-ph/0208008

\bibitem[{{Reichart} {et~al.}(2001){Reichart}, {Lamb}, {Fenimore},
  {Ramirez-Ruiz}, {Cline}, \& {Hurley}}]{rlf+01}
{Reichart}, D.~E., {Lamb}, D.~Q., {Fenimore}, E.~E., {Ramirez-Ruiz}, E.,
  {Cline}, T.~L., and {Hurley}, K. 2001, ApJ, 552, 57

\bibitem[{Rhoads(1997{\natexlab{a}})}]{rho97a}
Rhoads, J.~E. 1997{\natexlab{a}}.
\newblock IAU Circ. No. 6793

\bibitem[{Rhoads(1997{\natexlab{b}})}]{rho97b}
Rhoads, J.~E. 1997{\natexlab{b}}, ApJ ({\it Letters}), 487, L1

\bibitem[{Ricker {et~al.}(2002{\natexlab{a}})}]{rak+02a}
Ricker {\it et al.} 2002{\natexlab{a}}, {GCN} notice 1220

\bibitem[{Ricker {et~al.}(2002{\natexlab{b}})}]{rak+02b}
---. 2002{\natexlab{b}}, {GCN} notice 1315

\bibitem[{{Ricker} {et~al.}(2002){Ricker}, {Hurley}, {Lamb}, {Woosley},
  {Atteia}, {Kawai}, {Vanderspek}, {Crew}, {Doty}, {Villasenor}, {Prigozhin},
  {Monnelly}, {Butler}, {Matsuoka}, {Shirasaki}, {Tamagawa}, {Torii},
  {Sakamoto}, {Yoshida}, {Fenimore}, {Galassi}, {Tavenner}, {Donaghy},
  {Graziani}, {Boer}, {Dezalay}, {Niel}, {Olive}, {Vedrenne}, {Cline},
  {Jernigan}, {Levine}, {Martel}, {Morgan}, {Braga}, {Manchanda}, {Pizzichini},
  {Takagishi}, \& {Yamauchi}}]{rhl+02}
{Ricker}, G. {\it et al.} 2002, ApJ ({\it Letters}), 571, L127

\bibitem[{{Rossi} {et~al.}(2002){Rossi}, {Lazzati}, \& {Rees}}]{rlr02}
{Rossi}, E., {Lazzati}, D., and {Rees}, M.~J. 2002, MNRAS, 332, 945

\bibitem[{{Sagar} {et~al.}(2000){Sagar}, {Mohan}, {Pandey}, {Pandey}, \&
  {Castro-Tirado}}]{smp+00}
{Sagar}, R., {Mohan}, V., {Pandey}, A.~K., {Pandey}, S.~B., and
  {Castro-Tirado}, A.~J. 2000, Bulletin of the Astronomical Society of India,
  28, 15

\bibitem[{{Salmonson}(2000)}]{sal00}
{Salmonson}, J.~D. 2000, ApJ ({\it Letters}), 544, L115

\bibitem[{{Sari} {et~al.}(1999){Sari}, {Piran}, \& {Halpern}}]{sph99}
{Sari}, R., {Piran}, T., and {Halpern}, J.~P. 1999, ApJ ({\it Letters}), 519,
  L17

\bibitem[{{Sari} {et~al.}(1998){Sari}, {Piran}, \& {Narayan}}]{spn98}
{Sari}, R., {Piran}, T., and {Narayan}, R. 1998, ApJ ({\it Letters}), 497, L17

\bibitem[{Schaefer(2003)}]{sch03}
Schaefer, B. 2003, in press

\bibitem[{Schaefer {et~al.}(2002)}]{sgh+02}
Schaefer, B.~E. {\it et al.} 2002, {Submitted} to Ap.J; astro-ph/0211189

\bibitem[{Spergel {et~al.}(2003)}]{svp+03}
Spergel, D.~N. {\it et al.} 2003, apJ, submitted; astro-ph/0302209

\bibitem[{{Stanek} {et~al.}(1999){Stanek}, {Garnavich}, {Kaluzny}, {Pych}, \&
  {Thompson}}]{sgk+99b}
{Stanek}, K.~Z., {Garnavich}, P.~M., {Kaluzny}, J., {Pych}, W., and {Thompson},
  I. 1999, ApJ ({\it Letters}), 522, L39

\bibitem[{Stanek {et~al.}(2003)Stanek, Matheson, Garnavich, Martini, Berlind,
  Caldwell, Challis, Brown, Schild, Krisciunas, Calkins, Lee, Hathi, Jansen,
  Windhorst, Echevarria, Eisenstein, Pindor, Olszewski, Harding, Holland, \&
  Bersier}]{smg+03}
Stanek, K.~Z. {\it et al.} 2003, astro-ph/0304173

\bibitem[{{Tan} {et~al.}(2001){Tan}, {Matzner}, \& {McKee}}]{tmm01}
{Tan}, J.~C., {Matzner}, C.~D., and {McKee}, C.~F. 2001, ApJ, 551, 946

\bibitem[{{van den Bergh} \& {Pazder}(1992)}]{vp92}
{van den Bergh}, S. and {Pazder}, J. 1992, ApJ, 390, 34

\bibitem[{Vreeswijk {et~al.}(2002)}]{vfh+02}
Vreeswijk {\it et al.} 2002, {GCN} notice 1785

\bibitem[{{Vreeswijk} {et~al.}(2001){Vreeswijk}, {Fruchter}, {Kaper}, {Rol},
  {Galama}, {van Paradijs}, {Kouveliotou}, {Wijers}, {Pian}, {Palazzi},
  {Masetti}, {Frontera}, {Savaglio}, {Reinsch}, {Hessman}, {Beuermann},
  {Nicklas}, \& {van den Heuvel}}]{vfk+01}
{Vreeswijk}, P.~M. {\it et al.} 2001, ApJ, 546, 672

\bibitem[{Vreeswijk {et~al.}(1999)Vreeswijk, Rol, Hjorth, Kouveliotou, Pian,
  Palazzi, Pedersen, Gorosabel, Castro-Tirado, \& Greiner}]{vrh+99}
Vreeswijk, P.~M. {\it et al.} 1999, {GCN} notice 496

\bibitem[{Woods {et~al.}(1998)Woods, Kippen, \& Connaughton}]{wkc98}
Woods, P., Kippen, R.~M., and Connaughton, V. 1998, {GCN} notice 112

\bibitem[{{Yost} {et~al.}(2002){Yost}, {Frail}, {Harrison}, {Sari}, {Reichart},
  {Bloom}, {Kulkarni}, {Moriarty-Schieven}, {Djorgovski}, {Price}, {Goodrich},
  {Larkin}, {Walter}, {Shepherd}, {Fox}, {Taylor}, {Berger}, \&
  {Galama}}]{yfh+02}
{Yost}, S.~A. {\it et al.} 2002, ApJ, 577, 155

\bibitem[{{Zhang} \& {M{\' e}sz{\' a}ros}(2002)}]{zm02a}
{Zhang}, B. and {M{\' e}sz{\' a}ros}, P. 2002, ApJ, 571, 876

\end{thebibliography}

\newpage

\newpage

\newpage

\begin{deluxetable}{lccccccccc}
\tabletypesize{\scriptsize}
\tablecaption{
Observed and modeled data for GRB energy determination\label{tab:param}}
\tablecolumns{8}
\label{tab:param}
\tablehead{
\colhead{} & \colhead{} & \multicolumn{2}{c}{Fluence} \\
\colhead{GRB} & \colhead{$z$} & \colhead{$S_\gamma$ $\times 10^{-6}$} & 
\colhead{Bandpass\tablenotemark{a}} & \colhead{$k$\tablenotemark{b}} & \colhead{$t_{\rm jet}$} &
\colhead{$n$} & \colhead{Refs.} \\
\multicolumn{2}{c}{} & \colhead{erg cm$^{-2}$} & keV 
& \colhead{} & \colhead{day} & \colhead{cm$^{-3}$}}
\startdata
 970228 &  0.6950 & $ 11.70\pm{  2.00}$ &  1.5,   700 & $  0.830\pm{ 0.041}$ &                 \nodata &                 \nodata &            1, 2 \\*
 970508 &  0.8349 &                3.17 &   20,  2000 & $  0.814\pm{ 0.041}$ & $25.00\pm{ 5.00}$       & $  1.00$                &      3, 4, 5, 5 \\*
 970828 &  0.9578 &               96.00 &   20,  2000 & $  0.823\pm{ 0.036}$ & $ 2.20\pm{ 0.40}$       &                 \nodata &         6, 4, 6 \\*
 971214 &  3.4180 &                9.44 &   20,  2000 & $  0.804\pm{ 0.057}$ & $>   2.5$               &                 \nodata &         7, 4, 7 \\*
 980326 & \nodata &                0.92 &   20,  2000 &              \nodata & $<   0.4$               &                 \nodata &            4, 8 \\*
 980329 & \nodata &               55.10 &   20,  2000 &              \nodata & $<   1.0$               & $20.00\pm{10.00}$       &         4, 9, 9 \\*
 980425 &  0.0085 &                3.87 &   20,  2000 & $  1.002\pm{ 0.000}$ &                 \nodata &                 \nodata &           10, 4 \\*
 980519 & \nodata &               10.30 &   20,  2000 &              \nodata & $ 0.55\pm{ 0.17}$       & $ 0.14^{+0.32}_{-0.03}$ &       4, 11, 12 \\*
 980613 &  1.0969 & $  1.71\pm{  0.25}$ &   20,  2000 & $  0.863\pm{ 0.110}$ & $>   3.1$               &                 \nodata &      13, 14, 15 \\*
 980703 &  0.9662 &               22.60 &   20,  2000 & $  0.940\pm{ 0.041}$ & $ 3.40\pm{ 0.50}$       & $28.00\pm{10.00}$       &   16, 4, 17, 17 \\*
 981226 & \nodata & $  0.40\pm{  0.10}$ &   40,   700 &              \nodata & $>   5.0$               &                 \nodata &          18, 19 \\*
 990123 &  1.6004 &              268.00 &   20,  2000 & $  0.720\pm{ 0.052}$ & $ 2.04\pm{ 0.46}$       &                 \nodata &       20, 4, 20 \\*
 990506 &  1.3066 &              194.00 &   20,  2000 & $  0.873\pm{ 0.054}$ &                 \nodata &                 \nodata &           21, 4 \\*
 990510 &  1.6187 &               22.60 &   20,  2000 & $  1.026\pm{ 0.055}$ & $ 1.20\pm{ 0.08}$       & $ 0.29^{+0.11}_{-0.15}$ &   22, 4, 23, 12 \\*
 990705 &  0.8424 & $ 93.00\pm{  2.00}$ &   40,   700 & $  1.279\pm{ 0.098}$ & $ 1.00\pm{ 0.20}$       &                 \nodata &      24, 25, 26 \\*
 990712 &  0.4331 & $  6.50\pm{  0.30}$ &   40,   700 & $  1.387\pm{ 0.132}$ & $>  47.7$               &                 \nodata &      22, 27, 28 \\*
 991208 &  0.7055 &              100.00 &   25, 10000 & $  0.746\pm{ 0.206}$ & $<   2.1$               & $18.00^{+18.00}_{-6.00}$ &  29, 30, 31, 12 \\*
 991216 &  1.0200 &              194.00 &   20,  2000 & $  0.877\pm{ 0.042}$ & $ 1.20\pm{ 0.40}$       & $ 4.70^{+6.80}_{-1.80}$ &   32, 4, 33, 12 \\*
 000131 &  4.5110 &               41.80 &   20,  2000 & $  0.646\pm{ 0.074}$ & $<   3.5$               &                 \nodata &           34, 4 \\*
 000210 &  0.8463 & $ 61.00\pm{  2.00}$ &   40,   700 & $  1.278\pm{ 0.097}$ & $>   1.7$               &                 \nodata &      35, 35, 34 \\*
000301C &  2.0335 &                4.10 &   25,  1000 & $  0.928\pm{ 0.094}$ & $ 7.30\pm{ 0.50}$       & $27.00\pm{ 5.00}$       &  36, 37, 35, 12 \\*
 000418 &  1.1181 &               20.00 &   15,  1000 & $  0.997\pm{ 0.018}$ & $25.00\pm{ 5.00}$       & $27.00^{+250.00}_{-14.00}$ &  21, 38, 38, 12 \\*
 000630 & \nodata &                2.00 &   25,   100 &              \nodata & $>   4.0$               &                 \nodata &          30, 39 \\*
 000911 &  1.0585 &              230.00 &   15,  8000 & $  0.508\pm{ 0.063}$ & $<   1.5$               &                 \nodata &      40, 40, 40 \\*
 000926 &  2.0369 &                6.20 &   25,   100 & $  3.912\pm{ 1.328}$ & $ 1.80\pm{ 0.10}$       & $27.00\pm{ 3.00}$       &  41, 42, 43, 43 \\*
 010222 &  1.4768 & $120.00\pm{  3.00}$ &    2,   700 & $  1.115\pm{ 0.004}$ & $ 0.93^{+0.15}_{-0.06}$ & $  1.70$                &  44, 45, 46, 12 \\*
 010921 &  0.4509 & $ 15.40\pm{  0.20}$ &    8,   400 & $  1.475\pm{ 0.289}$ & $33.00\pm{ 6.50}$       &                 \nodata &      47, 48, 49 \\*
 011121 &  0.3620 &               24.00 &   25,   100 & $  4.996\pm{ 2.006}$ & $>   7.0$               &                 \nodata &      50, 51, 51 \\*
 011211 &  2.1400 &                5.00 &   40,   700 & $  1.068\pm{ 0.084}$ & $ 1.77\pm{ 0.28}$       &                 \nodata &      52, 53, 54 \\*
 020124 & \nodata &                3.00 &    8,    85 &              \nodata &                 \nodata &                 \nodata &              55 \\*
 020405 &  0.6899 & $ 38.00\pm{  4.00}$ &   50,   700 & $  1.318\pm{ 0.096}$ & $ 1.67\pm{ 0.52}$       &                 \nodata &          47, 56 \\*
 020331 & \nodata &                0.40 &    8,    40 &              \nodata &                 \nodata &                 \nodata &              57 \\*
 020813 &  1.2540 &               38.00 &   25,   100 & $  4.336\pm{ 1.682}$ & $ 0.43\pm{ 0.06}$       &                 \nodata &      58, 59, 58 \\*
 021004 &  2.3320 &                3.20 &    7,   400 & $  1.188\pm{ 0.098}$ & $ 7.60\pm{ 0.30}$       & $30.00^{+270.00}_{-27.00}$ &  60, 61, 62, 63 \\*
 021211 &  1.0060 &                1.00 &    8,    40 & $ 12.345\pm{ 6.462}$ &                 \nodata &                 \nodata &          64, 65  \enddata 

{}\vskip -0.05in\tablecomments{
References are given in order for the redshift, 
fluence, jet-break time ($t_{\rm jet}$), and density ($n$). Where
several redshift measurements are available for the same burst, we
choose the most precise determination. For fluence measurement, we
choose the measurement with the most precisely measured prompt burst
spectrum. For bursts with more than one $t_{\rm jet}$ or $n$
measurement, we choose the reference where the most broadband data
were used for afterglow modeling, preferring modeling that includes
radio afterglow measurements.}

\tablenotetext{a}{The energy range over which the fluence was reported.}

\tablenotetext{b}{The prompt burst spectral $k$-correction, 
as defined in \citet{bfs01}, used to transform the observed fluence
$S_\gamma$ in the particular bandpass to a standard restframe bandpass of
20--2000 keV.}

\vskip -0.1in
\tablerefs{ \tiny
1. \citet{bdk01}; 2. \citet{fac+98}; 3. \citet{bdkf98};
4. \citet{pjb00}; 5. \citet{fwk00}; 6. \citet{dfk+01};
7. \citet{kdr+98}; 8. \citet{ggv+98c}; 9. \citet{yfh+02};
10. \citet{gvvk+99}; 11. \citet{jhb+00}; 12. \citet{pk02};
13. \citet{dbk00}; 14. \citet{wkc98}; 15. \citet{hf98};
16. \citet{dkb+98b}; 17. \citet{fyb+03}; 18. \citet{faa+00};
19. \citet{fkb+99}; 20. \citet{kdo+99}; 21. \citet{bbk+03};
22. \citet{vfk+01}; 23. \citet{hbf+99}; see also \citet{sgk+99b}; 24. \citet{fdm+02};
25. \citet{afv+00}; 26. \citet{mpp+00}; 27. \citet{fro00};
28. \citet{fvhp00}; 29. \citet{csg+01}; 30. \citet{hcm+00};
31. \citet{smp+00}; 32. \citet{vrh+99}; 33. \citet{hum+00};
34. \citet{ahp+00}; 35. \citet{pfg+02}; 36. \citet{cdd+00};
37. \citet{bsf+00}; 38. \citet{bdf+01}; 39. \citet{fjg+01};
40. \citet{pbk+02}; 41. \citet{cdk+00}; 42. \citet{phg+01};
43. \citet{hys+01}; 44. \citet{mhk+02}; 45. \citet{zka+01};
46. \citet{grb+03}; see also \citet{jpg+01}; 47. \citet{pkb+02};
48. \citet{rhl+02}; 49. \citet{pks+02}; 50. \citet{gsw+03};
51. \citet{pbr+02}; 52. \citet{hsg+02}; 53. \citet{fag+02};
54. \citet{fox+02}; 55. \citet{rak+02a}; 56. \citet{pkb+03};
57. \citet{rak+02b}; 58. \citet{bsc+03}; 59. \citet{hcm+02};
60. \citet{mgf+02}; 61. \citet{lra+02}; 62. \citet{psr+02};
63. \citet{sgh+02}; 64. \citet{vfh+02}; 65. \citet{cvv+02} }

\end{deluxetable}

\newpage

\begin{deluxetable}{lccccccccc}
\tabletypesize{\footnotesize}
\tablecaption{Computed GRB Energetics and the Apparent GRB Distance Modulus\label{tab:dm}}
\tablecolumns{9}
\tablehead{
\colhead{} & \colhead{} & \colhead{} & \colhead{} &
\multicolumn{5}{c}{GRB Distance Modulus} \\
\colhead{GRB} & \colhead{$E_{\rm iso}(\gamma)$} & 
\colhead{$\theta_{\rm jet}$} & \colhead{$E_\gamma$\tablenotemark{a}} & 
\colhead{DM\tablenotemark{b}} & \multicolumn{4}{c}{Error Budget in DM\tablenotemark{c}} \\
\colhead{}   & \colhead{$\times 10^{50}$ erg} & 
\colhead{$^\circ$} & \colhead{$\times 10^{50}$ erg} &
\colhead{} & 
\colhead{$S_\gamma$} & \colhead{$k$} & \colhead{$n$} & \colhead{$t_{\rm jet}$}}
\startdata
 970228 & $  141.60\pm   25.19$ &               \nodata &               \nodata &               \nodata &  \nodata &  \nodata &  \nodata &  \nodata \\*
 970508 & $   54.57\pm    6.12$ & $    21.63\pm    1.67$ & $     3.84\pm    0.73$ & $    45.10\pm    0.47$ & 0.26 & 0.13 & 0.06 & 0.37 \\*
 970828 & $ 2198.20\pm  239.59$ & $     7.13\pm    0.50$ & $    17.00\pm    3.03$ & $    43.86\pm    0.45$ & 0.26 & 0.11 & 0.06 & 0.34 \\*
 971214 & $ 2105.36\pm  257.65$ & $>    5.54\pm    0.23$ & $>    9.84\pm    1.47$ & $<   47.83\pm    0.37$ & 0.26 & 0.18 & 0.06 & 0.19 \\*
 980613 & $   53.55\pm   10.39$ & $>   12.57\pm    0.58$ & $>    1.28\pm    0.28$ & $<   47.03\pm    0.54$ & 0.38 & 0.33 & 0.06 & 0.19 \\*
 980703 & $  601.20\pm   65.54$ & $    11.21\pm    0.81$ & $    11.47\pm    2.07$ & $    44.31\pm    0.45$ & 0.26 & 0.11 & 0.22 & 0.27 \\*
 990123 & $14378.92\pm 1778.12$ & $     4.93\pm    0.43$ & $    53.14\pm   11.33$ & $    44.00\pm    0.53$ & 0.26 & 0.19 & 0.06 & 0.42 \\*
 990506 & $ 8614.96\pm 1012.56$ &               \nodata &               \nodata &               \nodata &  \nodata &  \nodata &  \nodata &  \nodata \\*
 990510 & $ 1763.49\pm  200.18$ & $     3.36\pm    0.21$ & $     3.04\pm    0.51$ & $    47.13\pm    0.42$ & 0.26 & 0.14 & 0.28 & 0.12 \\*
 990705 & $ 2559.52\pm  203.18$ & $     5.33\pm    0.41$ & $    11.05\pm    1.91$ & $    43.98\pm    0.43$ & 0.06 & 0.20 & 0.06 & 0.37 \\*
 990712 & $   49.70\pm    5.27$ & $>   40.80\pm    1.70$ & $>   12.08\pm    1.60$ & $<   42.13\pm    0.33$ & 0.12 & 0.25 & 0.06 & 0.18 \\*
 991208 & $ 1121.55\pm  329.97$ & $<    8.64\pm    0.77$ & $<   12.72\pm    4.38$ & $>   43.35\pm    0.86$ & 0.26 & 0.71 & 0.36 & 0.19 \\*
 991216 & $ 5353.69\pm  593.92$ & $     4.57\pm    0.72$ & $    17.03\pm    5.65$ & $    44.02\pm    0.83$ & 0.26 & 0.12 & 0.46 & 0.62 \\*
 000131 & $11618.99\pm 1767.35$ & $<    4.68\pm    0.21$ & $<   38.66\pm    6.79$ & $>   47.06\pm    0.44$ & 0.26 & 0.30 & 0.06 & 0.19 \\*
 000210 & $ 1693.26\pm  140.51$ & $>    6.84\pm    0.28$ & $>   12.04\pm    1.40$ & $<   43.90\pm    0.29$ & 0.08 & 0.20 & 0.06 & 0.19 \\*
000301C & $  437.49\pm   62.21$ & $    13.14\pm    0.51$ & $    11.46\pm    1.86$ & $    46.30\pm    0.40$ & 0.26 & 0.26 & 0.12 & 0.13 \\*
 000418 & $  751.37\pm   76.31$ & $    22.30\pm    6.34$ & $    56.20\pm   32.06$ & $    42.97\pm    1.43$ & 0.26 & 0.05 & 1.35 & 0.37 \\*
 000911 & $ 3954.79\pm  630.19$ & $<    5.63\pm    0.25$ & $<   19.09\pm    3.48$ & $>   44.00\pm    0.46$ & 0.26 & 0.32 & 0.06 & 0.19 \\*
 000926 & $ 2797.36\pm  990.19$ & $     6.16\pm    0.31$ & $    16.17\pm    5.95$ & $    45.93\pm    0.92$ & 0.26 & 0.87 & 0.07 & 0.10 \\*
 010222 & $ 8578.41\pm  216.66$ & $     3.20\pm    0.13$ & $    13.33\pm    1.13$ & $    45.28\pm    0.21$ & 0.06 & 0.01 & 0.06 & 0.19 \\*
 010921 & $  136.11\pm   26.70$ & $    31.19\pm    2.46$ & $    19.67\pm    4.90$ & $    41.71\pm    0.62$ & 0.03 & 0.50 & 0.06 & 0.36 \\*
 011121 & $  455.59\pm  188.54$ & $>   15.35\pm    1.00$ & $>   16.26\pm    7.05$ & $<   41.35\pm    1.08$ & 0.26 & 1.03 & 0.06 & 0.19 \\*
 011211 & $  672.34\pm   85.68$ & $     6.38\pm    0.40$ & $     4.17\pm    0.74$ & $    47.54\pm    0.45$ & 0.26 & 0.20 & 0.06 & 0.30 \\*
 020405 & $  720.08\pm   92.23$ & $     7.81\pm    0.93$ & $     6.68\pm    1.80$ & $    43.99\pm    0.67$ & 0.27 & 0.19 & 0.06 & 0.58 \\*
 020813 & $ 7749.76\pm 3105.50$ & $     3.13\pm    0.23$ & $    11.58\pm    4.94$ & $    45.00\pm    1.07$ & 0.26 & 1.00 & 0.06 & 0.26 \\*
 021004 & $  556.01\pm   72.10$ & $    12.67\pm    4.51$ & $    13.53\pm    9.77$ & $    46.48\pm    1.80$ & 0.26 & 0.21 & 1.77 & 0.07 \\*
 021211 & $  378.24\pm  201.58$ &               \nodata &               \nodata &               \nodata &  \nodata &  \nodata &  \nodata &  \nodata  \enddata 

 \tablenotetext{a} {$E_\gamma$ is computed using equation
 \ref{eq:egamma}. The errors on the four observable parameters $k$,
 $t_{\rm jet}$, $S_\gamma$, and $n$ are taken from Table 1. When no
 errors are known for a given parameter we assume a fractional error
 of 10\%. We have taken $\Omega_m = 0.3$ and $\Omega_\Lambda = 0.7$
 for these energetics calculations.}

\tablenotetext{b}
{DM is apparent GRB distance modulus calculated from equation
\ref{eq:dm1}, assuming ${\cal E}_\gamma = 1.33$ foe. The cosmology as 
noted above was used for the calculating, but note that the
measurement on DM is relatively insensitive to the choice of
cosmological parameters (see footnote \ref{foot:cosmo}).}

\tablenotetext{c}
{These columns give the error contribution to the DM measurement from
each observable parameter. The error on DM is the quadrature sum of
these four numbers. Divide these numbers by 2.5 to get fractional
error contribution to $E_\gamma$.}

\end{deluxetable} \end{document}